\documentclass[preprintnumbers,aps,amsmath,amssymb,prl,twocolumn,showpacs,superscriptaddress]{revtex4-1}

\usepackage[dvipdfmx]{graphicx}
\usepackage{amsthm}
\usepackage{enumerate}
\usepackage{amsmath}
\usepackage{amssymb}
\usepackage{epstopdf}
\usepackage{color}
\usepackage{euscript}
\usepackage{float}
\usepackage{latexsym}
\usepackage{amsfonts}
\usepackage{bm} 
\usepackage{mathrsfs}
\usepackage{subfigure}

\newcommand{\bra}[1]{\langle {#1} |}
\newcommand{\ket}[1]{| {#1} \rangle}
\newcommand{\abs}[1]{\left| {#1} \right|}
\newcommand{\norm}[1]{\abs{\abs{#1}}}

\newcommand{\tr}[1]{\mathrm{Tr}\left[ {#1} \right]}
\newcommand{\partialtr}[2]{\mathrm{Tr}_{#1}\left[ {#2} \right]}

\newcommand{\blank}[0]{\hspace{3.5mm}}

\newtheorem{theo}{Theorem}
\newtheorem{lem}{Lemma}
\newtheorem{cor}{Corollary}

\begin{document}

\title{Quantum algorithm for universal implementation of projective measurement of energy}

\author{Shojun Nakayama}
\affiliation{Department of Physics, Graduate School of Science,
University of Tokyo, 7-3-1, Hongo, Bunkyo-ku, Tokyo, Japan}

\author{Akihito Soeda}
\affiliation{Department of Physics, Graduate School of Science,
University of Tokyo, 7-3-1, Hongo, Bunkyo-ku, Tokyo, Japan}
\affiliation{Centre for Quantum Technologies, National University of Singapore, Singapore}

\author{Mio Murao}
\affiliation{Department of Physics, Graduate School of Science,
University of Tokyo, 7-3-1, Hongo, Bunkyo-ku, Tokyo, Japan}
\affiliation{Institute for Nano Quantum Information Electronics,
University of Tokyo, 4-6-1, Komaba, Meguro-ku, Tokyo, Japan}

\begin{abstract}
A projective measurement of energy (PME) on a quantum system is a quantum measurement, determined by the Hamiltonian of the system.
PME protocols exist when the Hamiltonian is given in advance.
Unknown Hamiltonians can be identified by quantum tomography, but {the} time cost to achieve a given accuracy increases exponentially
with the size of the quantum system.
In this letter, we improve the time cost by adapting quantum phase estimation, an algorithm designed for computational problems, to measurements on physical systems.  We present a PME protocol without quantum tomography {for} Hamiltonians whose dimension and energy scale are given but otherwise unknown.
Our protocol implements a PME to arbitrary accuracy without any dimension dependence on its time cost.
We also show that another computational quantum algorithm may be used for efficient estimation of the energy scale.
These algorithms show that computational quantum algorithms have applications beyond their original context with suitable modifications.
\end{abstract}

\maketitle

\section{Introduction}Projective measurement of energy (PME) is a quantum counterpart of an ideal energy measurement in classical mechanics.
A PME on a given system sets the system to an energy eigenstate and returns the corresponding energy eigenvalue.
A PME alone has no effect on {a} system already in an energy eigenstate, thus can be used to confirm that the system remains
in the initial energy eigenstate by repeating the same PME and observing that the outcomes remain unchanged.
These properties make PME {suitable} for detecting small effects on a quantum system that is subject to an external influence
such as gravity wave\,\cite{QND} or thermal fluctuation\,\cite{FT1,FT2,FT3}.

\begin{figure}[t]
\vspace{-3mm}
\includegraphics[clip, width=0.87\columnwidth]{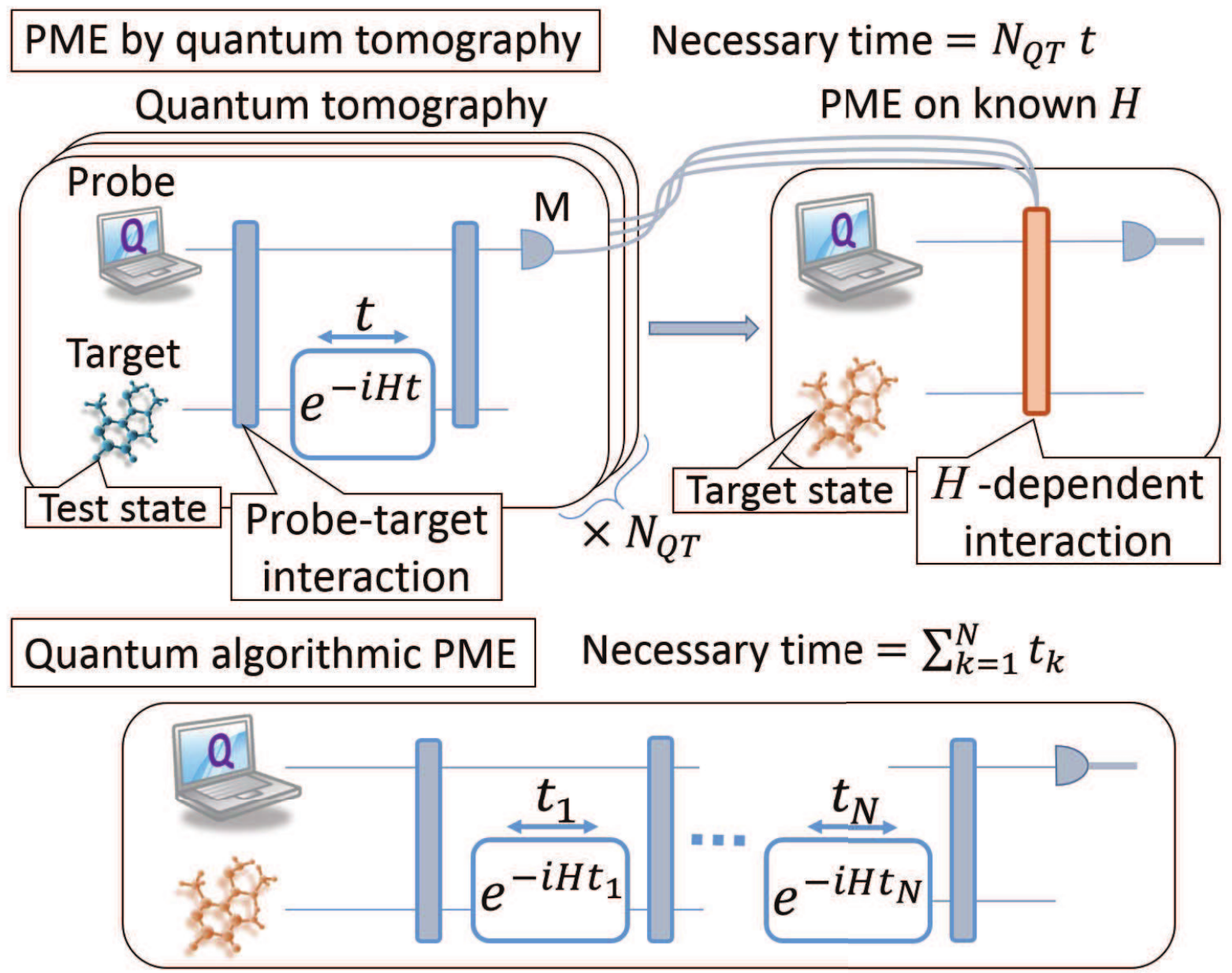}
\caption{(Color online) Schematic diagrams of PME protocols on a system of unknown self-Hamiltonian $H$ (labeled ``Target").  The blue boxes $\exp(-iHt)$ denote the target being let evolve for time $t$ with $\hbar\!=\!1$.  \textsf{M} is a quantum measurement which returns a numerical outcome. The implementation time is lower-bounded by the time required to induce the evolution of the target system, since there is no limit to the strength of the interaction induced from outside on the system in principle.  In the top protocol, $H$ is identified by quantum (process) tomography, with at least $N_{QT} = O(d^2)$ uses of the time evolution {$\exp(-iHt)$} for $d$-dimensional systems.  The quantum algorithmic PME (bottom) proposed in this letter avoids quantum tomography and all interactions are $H$-independent.
}\label{fig3png}
\end{figure}

In practice, a device that implements a quantum measurement must include a destructive component such as a photon detector.
PME being a nondestructive measurement {requires} another quantum system as a ``probe".
The system (commonly referred to as ``target") interacts with the probe,
and a direct measurement is performed only on the probe after the interaction (Fig.\,\ref{fig3png}).

An implementation protocol of PME is known for systems whose Hamiltonian $H$ is given in advance\,\cite{Aharonov}.
The protocol chooses the interaction between the probe and target according to $H$, so that the two quantum systems are appropriately entangled.
The entanglement assures that the measurement on the probe sets the target to an energy eigenstate, and that the outcome of the measurement identifies the respective energy eigenvalue.
The time needed to induce the entanglement can be made arbitrarily short by increasing the strength of the interaction.
Thus, PME of known $H$ can be implemented instantaneously in principle.

This protocol, however, does not take into account the time required to identify $H$.
Let us estimate the time cost by analyzing quantum process tomography\,\cite{Nielsen,tomo2} on the time evolution of the system.
{Process} tomography involves setting the target to various ``test states" and measuring the expectation value of appropriate observables for each resulting state after the time evolution.  A complete process tomography {for a system described by a $d$-dimensional Hilbert space $\mathcal{H}=\mathbb{C}^d$} {requires a number of observable, $O(d^2)$, equal to} the number of parameters in the Hamiltonian\,\cite{Baldwin}.

An accurate estimation of the expectation values needs to accumulate sufficient statistics.
Each use of the time evolution costs time $t$, hence the total time cost for the tomography to achieve a given accuracy for a $d$-dimensional system scales at least $O(d^2)$.
This implies that, if $H$ is unknown, the total implementation time for PME via process tomography grows at least exponentially in the number of subsystems {due to the exponential growth of the total dimension for composite systems}.

{Tomography} is required even if a PME is to be performed only once.
It extracts enough information to identify all the eigenspaces and eigenvalues of $H$, {so} the dimension dependence is unavoidable.
A single use of PME, however, does not reveal the exact description of the energy eigenspaces or the whole energy spectrum.
A more efficient PME protocol is needed.

To improve a PME protocol is to find a better quantum algorithm.
Some quantum algorithms are known to provide an efficient solution to computational problems\,\cite{NielsenChuang}.
These algorithms, however, assume that the dynamics of a quantum system can be ``switched off" at will, which does not hold in {this} problem.

In this paper, we introduce a more efficient PME protocol and show that we can remove the dimension-dependence in the time cost, for unknown Hamiltonians whose energy scale is given.
Our protocol exploits a modified version of {\it quantum phase estimation} ({QPE})\,\cite{Kitaev}.
Finally, we discuss an estimation protocol for the energy scale, based on an estimation of the trace of {a} unitary operator.
 We will show that another computational quantum algorithm, adapted from Ref.\,\cite{DQC1}, performs more efficiently than a complete tomography.

\section{Projective measurement by QPE} 
QPE is designed so that each run returns a good estimate for some eigenvalue of a given unitary operator $U=\sum_{k=1}^d \exp(i \theta_k) \ket{\theta_k}\bra{\theta_k}$ on $\mathcal{H}=\mathbb{C}^d$. 
Note that we assume $0\leq \theta_k < 2\pi$.
For a given input state $\ket{\theta_k}$, the corresponding phase $\theta_k$ is estimated by QPE.

An essential building block of QPE is a controlled-unitary operation $C_U$, which is a unitary gate that conditionally operates $U$ on a $d$-dimensional {\textit{target}} system {denoted by $\mathcal{H}_t=\mathbb{C}^d$} according to the state of an extra {\textit{control}} qubit  denoted by $\mathcal{H}_c=\mathbb{C}^2$.
Formally, the action of $C_U$ on $\mathcal{H}_c \otimes \mathcal{H}_t$ is defined by $C_U\ket{0}\ket{\varphi}\!=\!\ket{0}\ket{\varphi}$ and $C_U\ket{1}\ket{\varphi}\!=\!\ket{1}U\ket{\varphi}$ for any $\ket{\varphi} \in \mathcal{H}_t$ where $\{\ket{0}, \ket{1} \}$ forms the computational basis of $\mathcal{H}_c$.  To achieve $N$-bits estimation, QPE uses $N$ control qubits for applying $(C_U)^{2^{l-1}}$  between the $l$-th control qubit for each $l \in \{1,\dots ,N\}$ and the target.   We obtain an $N$-bit string $\{n_1, \cdots, n_{N}  \}$ of outcomes by the final measurements on the $N$ control qubits in the computational basis. {By} defining $\mathbf{n}_N:=\sum_{l=1}^N 2^{l-1} n_l$ and $f(\mathbf{n}_N) := \mathbf{n}_N/2^N$, the phase $\theta_k$ is estimated as $\theta_k = 2 \pi f(\mathbf{n}_N)$.

In the {limit} $N \rightarrow \infty$, $f(\mathbf{n}_N)$ can be regarded as a continuous variable $f$ with $0 \leq f \leq 1$.
For any $\theta_k$, {the probability $p_N(f(\mathbf{n}_N)|\theta_k)$ {to obtain} $\mathbf{n}_N$ {for an}  initial state $\ket{\theta_k}$} {approaches} the delta function $\delta(f - \theta_k / 2\pi)$ in distribution.
The distance between $p_N(f(\mathbf{n}_N)|\theta_k)$ and $\delta(f - \theta_k / 2\pi)$ is independent of $d$.
At the same limit, the target is transformed to an eigenstate by a projection onto the corresponding eigenspace induced by the final measurements of QPE.
Interested readers may refer to Appendix.~\ref{qpeap} for details of QPE.

\section{QPE and universal controllization}
The evolution of a target with Hamiltonian $H$ for time $t$ is given by the unitary operator $U(t)=\exp(-iHt)$, with $\hbar=1$.  It may appear that QP on $U(t)$ readily implements a projection onto the eigenspace corresponding to the estimated phase of  $U(t)$, which is also the desired PME of $H$ up to ambiguity due to the phase periodicity.
QPE assumes that $U$ is available in its quantum-controlled form, namely, $C_U$, but the time evolution operator is not.
Adding a quantum control to a quantum gate-- a task which we call {\it controllization}-- is not trivial when $U$ is unknown.
In this paper we introduce {\it universal controllization}, a quantum subroutine that approximately implements controllization for unknown $U$.

We introduce a $d$-dimensional ancillary system denoted by $\mathcal{H}_a=\mathbb{C}^d$ and define a unitary gate $W_U := C_S \left( {\mathbb I}_2 \otimes U \otimes {\mathbb I}_d \right) C_S$ on $\mathcal{H}_c \otimes \mathcal{H}_t \otimes \mathcal{H}_a$ where $C_S$ is a unitary gate called the controlled-swap operation defined by $C_S\ket{0}\ket{\psi}\ket{\phi}\!=\!\ket{0}\ket{\phi}\ket{\psi}$ and $C_S\ket{1}\ket{\psi}\ket{\phi}\!=\!\ket{1}\ket{\psi}\ket{\phi}$, for any $\ket{\psi}, \ket{\phi}\!\in\!{\mathbb C}^d$, and $\mathbb{I}_k$ denotes the $k \times k$ identity matrix.
We call $W_U$ a \textit{classically conditioned quantum gate} since it perfectly simulates $C_U$ when the control qubit is in a state $\ket{0}$ or $\ket{1}$.
But $W_U$ deviates from $C_U$ for a general input state $\ket{\eta} = \alpha \ket{0} + \beta \ket{1}$ in the control. 
Since $W_U \ket{\eta}\ket{\psi}\ket{\phi}\!=\!\alpha \ket{0} \ket{\psi} U \ket{\phi}\!+\!\beta \ket{1} U \ket{\psi} \ket{\phi}$, the ancilla system is also entangled to the control and target systems and thus decoherence occurs in the control-target system in general.
If we can prepare an eigenstate of $U$ {in} the ancilla system, exact implementations of $C_U$ is possible \cite{Friis,UnknownCont,Xiao}, but such implementations require knowledge on $U$.
Other know controllization schemes \,\cite{Barenco,Metropolis} also require that the quantum gate is at least partially known.

It is even proven that an exact controllization is impossible within quantum mechanics\,\cite{Mateus,Friis}.  These results are derived assuming that the input quantum gate is a blackbox.
The unitary operator $U(t)$, on the other hand, has a tunable parameter, namely, the evolution duration $t$.  We exploit this feature and a decoupling method\,\cite{decoupling} used in quantum information theory to asymptotically implement a \textit{universal} controllization of $U(t)$.  The implementation accuracy of our controllization depends on the maximum difference between any two eigenvalues of $H$.

To reduce the decoherence by $W_U$, we need to make the resulting state of the ancilla depend as little as possible on the initial control-target state.
Let us prepare the ancilla in the completely mixed state ${\mathbb I}_d/d$, so that the state of the ancilla remains the same at least when the control qubit is in $\ket{0}$ or $\ket{1}$ for any given $U$.
We consider the reduced map on the control-target system,
\begin{equation}
\Gamma_{U} \left[ \rho \right] :=\partialtr{{\mathcal H}_a}{W_{U}( \rho \otimes {\mathbb I}_d/d)W_{U}^\dagger },
\end{equation}
where $\rho$ is a density matrix on $\mathbb{C}^2 \otimes \mathcal{H}_t$.
We call the map $\Gamma_{U}$ as {\it pseudo controllization}.
For $\rho = \ket{\eta}\bra{\eta} \otimes \ket{\psi}\bra{\psi}$, we have
\begin{multline}
 \Gamma_U\left[\ket{\eta}\bra{\eta} \otimes \ket{\psi}\bra{\psi} \right] = C_U \left( \ket{\eta}\bra{\eta} \otimes \ket{\psi}\bra{\psi} \right) C_U^\dag + \\
 \left[ \alpha \beta^* \ket{0}\bra{1} \otimes \ket{\psi}\bra{\psi} \left( \gamma_U -1 \right) U^\dag + c.c.\right], \label{diff}
\end{multline}
where $\gamma_U = \tr{U}/d$.
The second term in Eq.\,\eqref{diff} acts as a kind of phase damping noise on the control-target system.
The factor $\gamma_U - 1$ determines the deviation of the reduced map $\Gamma_U$ from the ideal controllization.
We define the \textit{coherence factor} $a_U := |\gamma_U|$ and a phase factor $e^{i\varphi_U} := \gamma_U/|\gamma_U|$.  Notice that $1-a_U \leq \left|\gamma_U - 1\right|$.  Thus, the phase damping noise is {minimized}
if we regard $\Gamma_U$ as an approximation of $C_{U'}$ for $U' = e^{-i \varphi_U}U$.
In a sense, $\Gamma_U$ implements a noisy controlled-unitary operation, where the magnitude of the noise is determined by a positive quantity $1\!-\!a_U$.

We further reduce the dependence of the ancilla on the initial control-target state by use of a set $\{\sigma_r\}$ of unitary operations on the ancilla such that
\begin{equation} \label{refresh}
 \frac{1}{d^2} \sum_r  \sigma_r W_{U}(\rho_{\rm tot} \otimes {\mathbb I}_d/d)W_{U}^\dagger \sigma_r^\dagger = \Gamma_{U} \left[ \rho_{\rm tot}\right] \otimes {\mathbb I}_d/d.
\end{equation}
Note that the ancilla is  ``refreshed" to the completely mixed state only by operations on the aniclla.  (Such a random operation has been extensively applied to questions in quantum communication\,\cite{decoupling}.)
We divide $W_{U(t)}$ into $m$ repetitions of $W_{U(t/m)}$, each followed by the refreshing operation\,\eqref{refresh}.  Here, $m$ fixes the refresh rate.
The strength of the noise after each refreshing operation is $O(\frac{1}{m^2})$.
Thus the total effect of the noise scales $O(m\!\times\!\frac{1}{m^2}) = O(\frac{1}{m})$, which vanishes in the asymptotic limit of $m \rightarrow \infty$~{.} (see Appendix.~\ref{concentphi}, for details).
This phenomenon is mathematically analogous to the quantum Zeno effect\,\cite{Zeno}.

We call this asymptotic implementation of a controlled-unitary operation including the repeated refreshing operation, {\it universal controllization}.
For {finite} $m$, the universal controllization approximates the controlled-unitary operation $C_{U^{[m]}(t)}$, where $U^{[m]}(t)\!=\!e^{-im\varphi_{U(t/m)}}U(t)$.  With $m\!\rightarrow\!\infty$, $\exp(im\varphi_{U(t/m)})$ converges to $\exp(-i\tr{H}t/d)$.  In a sense, {universal controllization} fixes the reference point of the energy of $H$ so that $\tr{H} = 0$.
A more detailed discussion of universal controllization is presented in Appendix.~\ref{universal}.

\section{PME by universal controllization}A perfect PME for a system with a Hamiltonian $H$ is distinguished from other quantum operations by two properties.
First, the system remains in the same eigenstate when a PME is applied consecutively.  Second, the outcomes of the consecutive measurements are all precisely equal to $E_k$.  The probability density $p(E|E_k)$ of obtaining $E$ as the outcome must be the delta function $\delta(E;E_k) := \delta(E - E_k)$.
Conversely, the only measurement satisfying these properties is a perfect PME.

A subtlety is that a perfect PME for $H$ and for $H - \lambda \mathbb{I}$ should be considered equivalent, since two Hamiltonians with different reference points of energy are physically equivalent.  A measurement scheme is regarded as a perfect PME for $H$ if $p(E|E_k) = \delta(E;E_k-\lambda)$ as long as $\lambda$ is independent of $k$.

Our PME protocol uses {QPE} on the time evolution operator $U(t)$ with $C_{U^{[m]}(t)}$ implemented by {universal controllization}.  Here, the control qubits and ancilla of the universal controllization serve as the probe.
The probe-target interaction is used to perform $W_{U(t/m)}$, the refreshing operations, and QFT.
The lower figure in Fig.\,\ref{fig3png} provides a conceptual diagram.

In the ideal case of $m \rightarrow \infty$ and $N \rightarrow \infty$, the modified {QPE} implements the projective measurement defined by the spectral decomposition of $\tilde{U}(t) = \exp \big(-i\tilde{H}t \big)$, where $\tilde{H} := H - \tr{H}\mathbb{I}$.  The outcome $f$ gives $-\tilde{E}_k t\,\pmod{2\pi}$ for some energy eigenvalue $\tilde{E}_k$ of $\tilde{H}$.

$\tilde{E}_k$ cannot be uniquely determined from $f$ for {general} $t$ due to the periodicity of the phase function $\exp(i \theta)$.
Let us restrict $t$ so that $\tilde{E}_k \in (\pi/t,-\pi/t)$, {namely,}
\begin{equation} \label{DelMax}
 \Delta_\mathrm{max} t \leq \pi/2,
\end{equation}
where $\Delta_\mathrm{max} = \max_{k,l} \big|\tilde{E}_k - \tilde{E}_l\big|$.
The energy eigenvalues are uniquely determined by
\begin{equation} \label{E}
 E[f] = \begin{cases} -2\pi f/t & f \in \big[0,\frac{1}{2}\big)\\
 -(2\pi f\!-\!2\pi)/t & f \in \big[\frac{1}{2},1\big) \end{cases}.
\end{equation}
Recall that the probability distribution of $f$ is the delta function $\delta(f\!-\!\theta_k/2\pi)$.
Thus, $p(E|E_k) = \delta(E;E_k\!-\!\tr{H})$, which is the desired function.  The projection onto the corresponding energy eigenspace is already guaranteed by {QPE}.

For {finite} $m$ and $N$, we continue to choose $t$ according to Eq.\,\eqref{DelMax} and estimate $E_k$ by Eq.\,\eqref{E} with $f$ replaced by $f(\mathbf{n}_N)$.
The implemented measurement is an approximation of a PME.
{A} target initially in an energy eigenstate $\ket{E_k}$ results in the same state at the end of the scheme.
One of the conditions for a perfect PME is still satisfied.
Thus, the accuracy of the scheme is determined by how close $p(E|E_k)$ for each $\ket{E_k}$ simulates a delta function $\delta(E;E_k-\lambda)$.

\section{Implementation accuracy and time cost} Recall that $p_N(f(\mathbf{n}_N)|\theta_k)$ in QPE needs to approach the delta function $\delta(f - \theta_k/2\pi)$ in $N \rightarrow \infty$ to achieve the projective measurement determined by $U$.  If each $C_U$ in QPE is replaced by the adapted classical controllization (i.e., substituted by $W_U$ and an ancilla), $p_N(f(\mathbf{n}_N)|\theta_k)$ does not converge to the delta function unless $ a_{U} = 1$.

\begin{figure}[t]
\begin{center}
\subfigure[$p^{[m]}_N(f(\mathbf{n}_N)|E_k)$ for a fixed $m$]{\hspace{6mm}
\includegraphics[width= 0.9\columnwidth,bb=0 0 900 430]{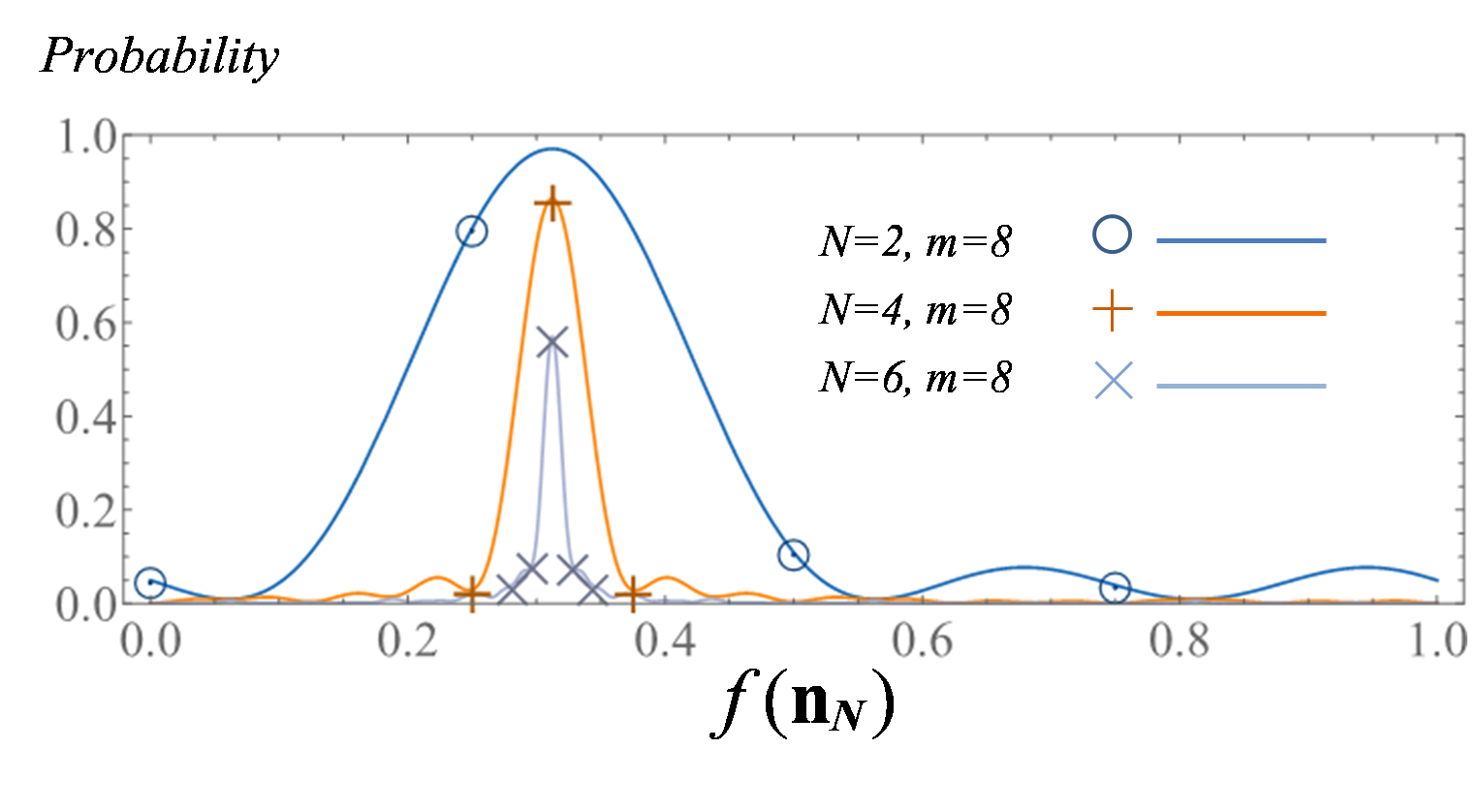}}\label{a}
\subfigure[$p^{[m]}_N(f(\mathbf{n}_N)|E_k)$ for an adaptively chosen $m$]{\hspace{6mm}
\includegraphics[width= 0.9\columnwidth,bb=0 0 900 430]{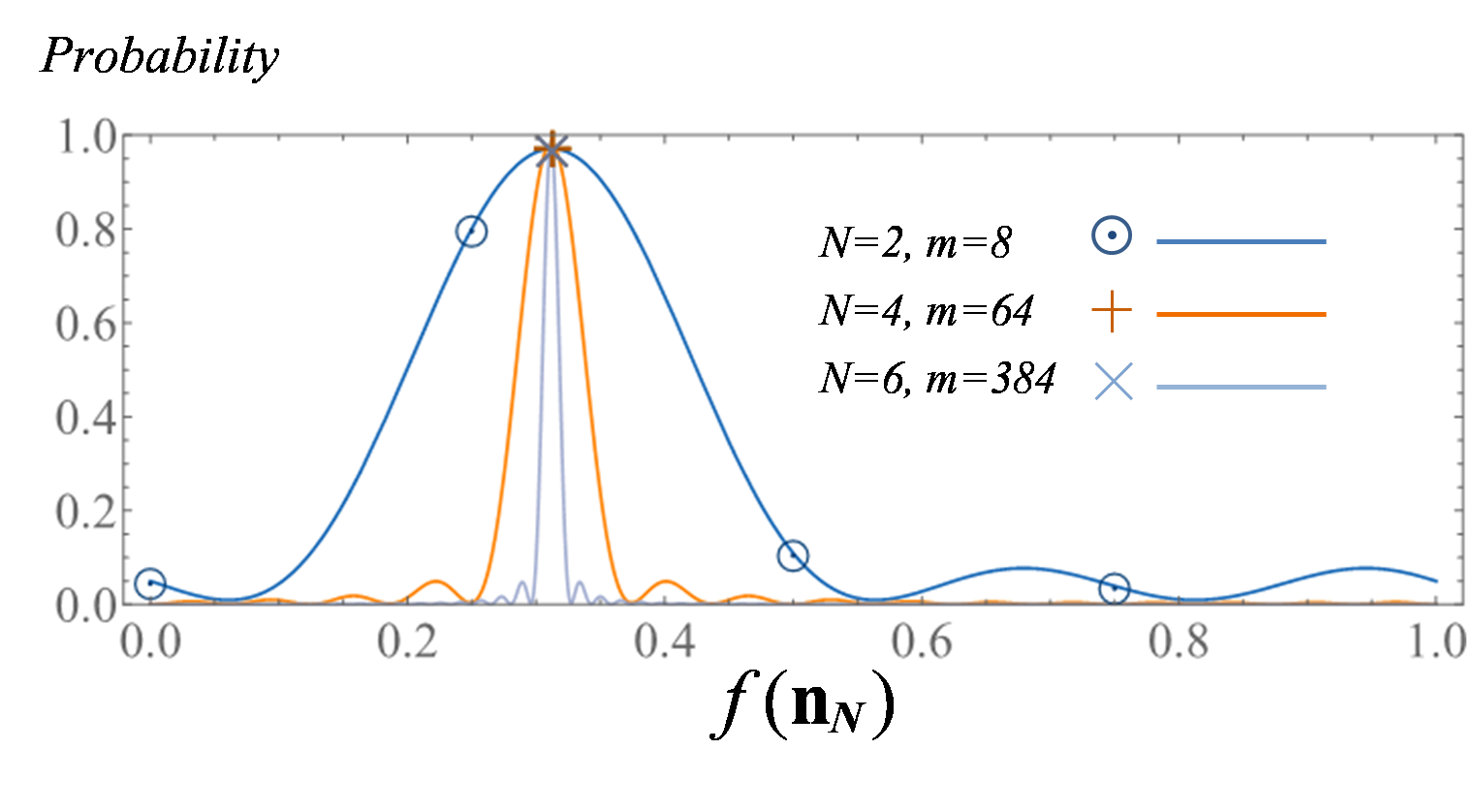}}
\caption{(Color online) Plots of probability distributions $p^{[m]}_N(f(\mathbf{n}_N)|E_k)$ and their envelope functions for target Hamiltonian $H\!=\!-\sum_{\lambda=0}^3\ket{E_0^{(\lambda)}}\bra{E^{(\lambda)}_0}\!+ \ket{E_1}\bra{E_1}$, $t\!=\!0.225\!\cdot\!\pi$, and setting $E_k\!=\!E_1$.
Each marker represents $p^{[m]}_N(f(\mathbf{n}_N)|E_k)$ {of} finding each outcome by a single round of {the} PME scheme. 
Fig.2 (a) presents $p^{[m]}_N(f(\mathbf{n}_N)|E_k)$ for $N=2,4,6$ for a fixed refreshing rate $m=8$.  Fig.2 (b) presents $p^{[m]}_N(f(\mathbf{n}_N)|E_k)$ for $N=2,4,6$ where each value of $m$ is adoptively chosen as a smallest integer satisfying $m \geq (\Delta_\mathrm{max} t)^2 N 2^{N-3} / \epsilon$ {and target error $\epsilon$ is set} to $0.25$.  In all cases, {markers corresponding to} probability less than one tenth of the target error ($0.025$) are omitted for clarity.  }\label{plots}
\end{center}
\end{figure}

Let us denote by $p^{[m]}_N(f(\mathbf{n}_N)|E_k)$ the probability distribution of $f(\mathbf{n}_N)$ for a given $m$, $N$, and initial state $\ket{E_k}$.
For a finite $m$, the universal controllization approximately controllizes $U' = \exp(-i(Ht\!-\!m\varphi_{U(t/m)}\mathbb{I}))$.  In this case, each run of the approximated {QPE} provides an estimate for the eigenvalue corresponding to $\ket{E_k}$, which is $\theta'_k\!=\!-E_k t\!+\!m\varphi_{U(t/m)}\pmod{2\pi}$.  
When $N$ increases, the deviation of $p^{[m]}_N(f(\mathbf{n}_N)|E_k)$ from $p_N(f(\mathbf{n}_N)|\theta'_k)$ caused by the controllization error prevents the function converging to a delta function.  (See Fig.~\ref{plots}.a)
The deviation can be bounded by $\left|p^{[m]}_N(f(\mathbf{n}_N)|E_k) - p_N(f(\mathbf{n}_N)|\theta'_k)\right|\!\leq\!\epsilon$ for any $\epsilon\!>\!0$ when $m$ is set to 
\begin{equation}
m \geq (\Delta_\mathrm{max} t)^2 N 2^{N-3} / \epsilon
\end{equation}
as  shown in Appendix.~\ref{AAqpe}.(See Fig.~\ref{plots} b for {examples}.) 

For a given refresh rate $m$, each universal controllization makes $m$ uses of $W_{U(t/m)}$, where the total evolution duration $(t/m)\!\times\!m\!=\!t$ is independent of $m$.
Hence, $p^{[m]}_N(f(\mathbf{n}_N)|E_k)$ can be brought arbitrarily close to $p_N(f(\mathbf{n}_N)|\theta'_k)$ without increasing the time cost.  The distribution $p_N(f(\mathbf{n}_N)|\theta'_k)$ is not a delta function for any finite $N$ even with perfect controlled-unitary operations $C_U$ (i.e., infinite $m$).  The cost doubles for each control qubit added, but the distance between $p_N(f(\mathbf{n}_N)|\theta'_k)$ and the delta function $\delta(E;\tilde{E}_k)$ is independent of the dimension of the target.  Hence, the implementation accuracy of PME can be improved without any dimension dependence.

\section{Quantum algorithmic estimation of {the energy scale}}We showed the existence of our PME protocol under the assumption that $\Delta_\mathrm{max}$ is known.
The assumption can be relaxed to {knowing} an \textit{upper bound} on $\Delta_\mathrm{max}$.
The bound may be estimated by quantum (process) tomography, but the tomography requires that a prior distribution of $H$ is given.
For a certain prior distribution, it is possible to estimate the bound by measuring the coherence factor $a_U$.
We observe that $a_U$ approaches $1$ as the product $\Delta_\mathrm{max} t$ decreases to $0$.
Thus, when $a_U$ is estimated to be close to 1,  it is possible that $\Delta_\mathrm{max} t$ is sufficiently small.
While this is not true for some Hamiltonians, the probability of such ``error" decreases exponentially in the dimension $d$ of the target for a particular class of prior distribution
(see Appendix.\,\ref{ES}).
Hence, we can reliably {estimate $a_U$.}

To estimate $a_U$, we modify the quantum algorithm presented in Ref.\,\cite{DQC1}.
The original algorithm outputs the trace $\tr{U}$ of an input unitary $U$, provided that the corresponding $C_U$ is available.
In our problem, we replace $C_U$ with $W_U$.
With this modification, the original algorithm returns $|\tr{U}|^2$ (See Appendix. \ref{esta}, for details), thus we obtain $a_U$ since $a_U^2\!=\!|\tr{U}|^2 / d^2$.
Clearly, this modified algorithm estimates $a_U$ much more efficiently than process tomography.

{\it Conclusion.---}In this letter, we presented an implementation protocol for a projective measurement of energy on a system driven by an unknown Hamiltonian with a given energy scale.
The implementation time cost of the protocol is independent of the dimension of the system unlike the one based on quantum process tomography.
The protocol is based on a computational quantum algorithm called {quantum phase estimation (QPE).}
We introduced universal controllization to make the computational algorithm executable without stopping the evolution of the target system.
Another computational quantum algorithm is shown to be effective in estimating the energy scale with a suitable modification.
This motivates {the} search for further applications of quantum algorithms outside their original computational context.

{\it Acknowledgments:} The authors thank T.~Sugiyama and H.~Nishimura for their insights and expertise.
This work is supported by the Project for Developing Innovation Systems of MEXT, Japan, the Global COE Program of MEXT Japan, and JSPS KAKENHI (Grant No.\,23540463,
No.\,23240001, and No.\,26330006).
The authors also gratefully acknowledge the ELC project (Grant-in-Aid for Scientific Research on Innovative Areas MEXT KAKENHI (Grant No.\,24106009)) for encouraging the research presented in this paper.
After the completion of our work, we were notified that our algorithm for calculating $|\tr{U(t)}|$ based on DQC1 has been
independently discovered by J.~Thompson, M.~Gu, K.~Modi, and V.~Vedral in ``Quantum Computing with Black-box Subroutines"\,\cite{Mile}.
We thank these authors for drawing their work to our attention.

\appendix
\def\thesection{\Alph{section}}
\section{Elements of quantum phase estimation}\label{qpeap}
This section provides details of {\it quantum phase estimation} (QPE) described in the second and the third sections in the main article on {\it Projective measurement by QPE} and {\it QPE and universal controllization}. 
QPE plays {a} crucial role in our protocol for {\it projective measurement of energy} (PME).  
A quantum circuit representation of the algorithm of QPE, the probability distribution of outcomes and the transformed state corresponding to each outcome by QPE are presented in Sec.~\ref{pdqpe}.  
In the main paper, we referred that the probability distribution of the outcome converges to a delta function at the limit where the number of control qubits goes infinity.  In Sec.~\ref{convtodelta}, we give the mathematical formulation of the statement.

\subsection{Probability distribution of outcomes and state change induced by {QPE}}\label{pdqpe}
QPE is designed so that each run returns a good estimate for one of the eigenvalues of a given unitary operator $U$.
QPE (originally proposed in \cite{Kitaev}) is usually described in the state-vector formalism.
In this subsection, we provide another description based on the density-matrix formalism to facilitate the comparison with the approximate QPE using universal controllization presented in Appendix.\,\ref{probout}.

\begin{figure}[!t]
 \begin{center}
  \includegraphics[clip, width=1\columnwidth]{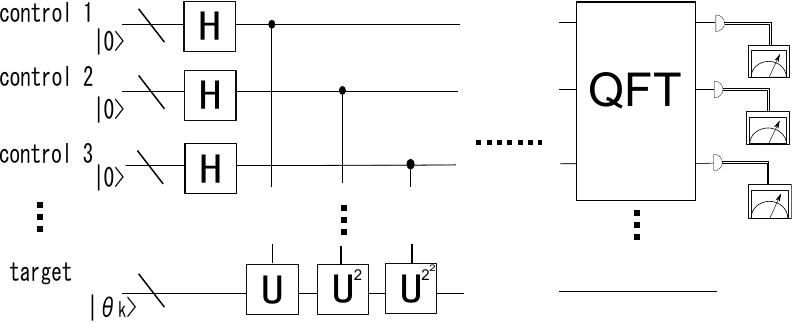}
 \end{center}
 \caption{A quantum circuit representation of QPE. The box QFT denotes the quantum Fourier transformation.  The final measurement is performed in the computational basis.
 }\label{PhaseEstimation}
\end{figure}

The circuit representation of QPE is given in Fig.\,\ref{PhaseEstimation}. 
Consider a target system ${\mathbb C}^d$ and a control system consisting of $N$-qubit systems ${\mathbb C}^{2^N}$. We set a basis of a qubit system $\mathbb{C}^2$ and denote the basis by $\{\ket{0}, \ket{1} \}$.

First, we initialize the control and target system as
\begin{equation}
 \ket{0 \dots 0} \bra{0 \dots 0} \otimes \ket{\theta_k} \bra{\theta_k},
\end{equation}
on $\mathbb{C}^{2^N} \otimes {\mathbb C}^d$, where $\ket{0 \dots 0} := \ket{0} \otimes \cdots \otimes \ket{0}$ and
$\ket{\theta_k}$ is an eigenvector corresponding to eigenvalue $e^{i \theta_k}$ of  $U$. 
The Hadamard gate $\mathrm{H}$ is then applied to each control qubit.
Note that $\mathrm{H}$ achieves 
\begin{align}
 \mathrm{H} \ket{0} &= (\ket{0} + \ket{1})/\sqrt{2},\\
 \mathrm{H} \ket{1}  &= (\ket{0} - \ket{1})/\sqrt{2}.
\end{align}

The state after this operation is given by
\begin{equation}
 \frac{1}{2^N}\sum_{\begin{subarray}aa_1,a_2,\dots a_N \\ b_1, b_2 ,\dots, b_N\end{subarray}} \ket{a_1a_2 \dots a_N}\bra{b_1 b_2 \dots b_N} 
	\otimes \ket{\theta_k}\bra{\theta_k},
\end{equation}
where $a_l, b_l \in \{0,1 \}$. 

A controlled-unitary operation $C_U$ of an unitary operation $U$ is defined as
\begin{eqnarray}
C_U := \ket{0}\bra{0} \otimes \mathbb{I}_d+\ket{1}\bra{1} \otimes U
\label{CU}
\end{eqnarray}
on ${\mathbb C}^2 \otimes {\mathbb C}^d$.
Here, $\mathbb{I}_d$ denotes the $d \times d$ identity matrix.
The superoperator representation $\mathcal{C}_U$, corresponding to $C_U$, is defined as
\begin{equation}
 \mathcal{C}_U \left[\rho\right] := C_U \rho C_U^\dagger.
\end{equation}
We choose the $l$-th control qubit and the target system and apply the controlled-unitary operation $\mathcal{C}_{U^{2^{l-1}}}$ for all $1 \leq l \leq N$. 
This transforms the state to
\begin{multline}
 \frac{1}{2^N}\sum_{\begin{subarray} aa_1,a_2,\dots a_N \\ b_1, b_2 ,\dots, b_N\end{subarray}} 
	\prod_{l=1}^N \exp\left( i2^{l-1} (a_l - b_l) \theta_k \right) \\
	\times \ket{a_1 a_2 \dots a_N}\bra{b_1 b_2 \dots b_N} 
	\otimes \ket{\theta_k}\bra{\theta_k}.\label{aftersecond}
\end{multline}

Finally, the quantum Fourier transformation is applied and {then} the control qubits are measured in the computational basis 
\begin{equation}
\left\{\ket{\mathbf{n}_N}=\ket{n_1n_2\dots n_N} \Big\vert \mathbf{n}_N=\sum_{l=1}^N n_l \cdot 2^{l-1} \right\}.
\end{equation}
The Fourier transformation and the measurement in the computational basis together are equivalent to performing a projective measurement on the state~\eqref{aftersecond} in the Fourier basis, i.e.,
\begin{equation}
\left \{\ket{f(\mathbf{n}_N)} \vert 0 \leq \mathbf{n}_N < 2^N  \right \},
\end{equation}
where $f(\mathbf{n}_N) := \mathbf{n}_N/2^N$ and 
\begin{equation}
 \ket{f(\mathbf{n}_N)} := \sum_{\mathbf{n}'_N=0}^{2^N-1} \frac{\exp \left({i 2 \pi f(\mathbf{n}_N) \mathbf{n}'_N }\right)}{\sqrt{2^{N}}} \ket{\mathbf{n}'_N}.\label{fourierbasis}
\end{equation}
The probability distribution $p_N(f(\mathbf{n}_N)\vert \theta_k)$ of obtaining the state $\ket{f(\mathbf{n}_N)}$ is calculated as 
\begin{align}
  &p_N(f(\mathbf{n}_N) \vert \theta_k) \nonumber\\
 &= \frac{1}{2^{2N}}\prod_{l=1}^N  \sum_{a,b \in \{0,1\}} \exp \big[ i2^{l-1} (a - b) (\theta_k - 2\pi f(\mathbf{n}_N))\big] \nonumber \\
 &= \frac{1}{2^N} \prod_{l=1}^N  \left[ 1 + \cos \big[ 2^{l-1} (\theta_k - 2\pi f(\mathbf{n}_N)) \big] \right].\label{prePN}
\end{align}
This simplifies to
\begin{equation}
 p_{N}(f(\mathbf{n}_N) \vert \theta_k) 
= \left( \frac{\sin \big[ 2^{N} \left( \theta_k - 2\pi f(\mathbf{n}_N) \right)/2\big]}{2^N \sin \big[ \left( \theta_k - 2 \pi f(\mathbf{n}_N) \right)/2 \big]}\right)^2, \label{PN}
\end{equation}
using
\begin{equation}
 1 + \cos(2^{l-1}x) = \frac{1}{2} \left( \frac{\sin(2^{l-1}x)}{\sin(2^{l-2}x)} \right)^2
\label{formuli}
\end{equation}
to Eq.\,\eqref{prePN}.

If we apply QPE to an arbitrarily superposed input state $\ket{\phi}=\sum_k \alpha_k \ket{\theta_k}$, where $\sum_k |\alpha_k|^2=1$, the probability distribution $p_N(f(\mathbf{n}_N)|
\phi)$ of obtaining the outcomes $\{n_1, \cdots, n_{N}  \}$ represented  in terms of $f(\mathbf{n}_N)$ is  given by
\begin{equation}
p_N(f(\mathbf{n}_N) | \phi) = \sum_k |\alpha_k|^2 p_N ( f(\mathbf{n}_N) | \theta_k).
\end{equation}
For the given outcome $f(\mathbf{n}_N)$, the corresponding output state of the target system can be calculated as
\begin{equation}
\ket{\phi^\prime_{f(\mathbf{n}_N)}} = \sum_k \alpha_k \sqrt{\frac{p_N (f(\mathbf{n}_N) | \theta_k)}{p_N(f(\mathbf{n}_N) | \phi) }} e^{iG(\theta_i,f(\mathbf{n}_N))}\ket{\theta_k}, \label{aftermeasurement}
\end{equation}
where \[ G(\theta_k,f(\mathbf{n}_N))=\left(2^N -1\right)\left(\theta_k - 2\pi f(\mathbf{n}_N)\right)/2. \]
Thus if the distribution $p_N(f(\mathbf{n}_N)\vert \theta_k)$ converges to the delta function $\delta(f-\theta_k/2\pi)$ for $N \rightarrow \infty$, then the output state $\ket{\phi^\prime_{f(\mathbf{n}_N)}}$ converges to a particular eigenstate $\ket{\theta_k}$.
In other words, in the limit of $N \rightarrow \infty$, we only obtain $f=\theta_k / 2 \pi$ with the target system in the corresponding eigenstate.
We see that QPE implements a projective measurement in the eigenbasis of $U$ for $N \rightarrow \infty$.
If the unitary $U$ is generated by a Hamiltonian $H$ as $U(t)= \exp \left( -iHt \right)$, QPE implements projective measurement of energy (PME) of $H$ up to ambiguity due to the phase periodicity.

\subsection{Convergence of $p_N(f(\mathbf{n}_N)\vert \theta_k)$ to a delta function}\label{convtodelta}
We {assumed} that $p_N(f(\mathbf{n}_N)\vert \theta_k)$ converges to the delta function $\delta(f-\theta_k/2\pi)$ for $N \rightarrow \infty$ in the last subsection.  
For each $\theta_k$ and any finite $N$, $f(\mathbf{n}_N)$ is a discrete random variable over $\big\{\mathbf{n}_N/2^N|\mathbf{n}_N=0,\dots,2^N-1\big\}$, distributed according to $p_{N}(f(\mathbf{n}_N) \vert \theta_k)$.  In contrast, $f$ is a \textit{continuous} random variable over real numbers $x$ in $0 \leq x \leq 1$.  
In the followings, we introduce a precise statement of the convergence to justify the assumption.
The convergence of a discrete random variable to a continuous one can be formulated with \textit{distribution functions}\,\cite{Billingsley}.

Let $[a,b]$ denote the set of real numbers $x$ such that $a \leq x \leq b$.
For any $A \subset [0,1]$, we define $\mu_N(A)$ by 
\begin{equation}
 \mu_N(A) = \sum_{\{\mathbf{n}_N|f(\mathbf{n}_N) \in A\}} p_{N}(f(\mathbf{n}_N) \vert \theta_k). 
 \label{deltameasure}
\end{equation}
If $A=[a,b]$ for $0 \leq a \leq b < \theta_k/2\pi$, we can bound $\mu_N(A)$ as
\begin{equation}
 \mu_N(A) \leq  N_A \left(\frac{1}{2^{N}\sin[(\theta_k-2\pi b)/2]}\right)^2, \label{pebound}
\end{equation}
where $N_A$ is the number of $f(\mathbf{n}_N)$ satisfying $f(\mathbf{n}_N) \in A$. 
Since \[N_A \leq 2^{N} (b-a) + 1,\]
we have
\begin{equation}
 \mu_N(A) \leq \frac{1}{\sin^2[(\theta_k - 2 \pi b)/2]} \frac{2^N(b-a)+1}{2^{2N}}. \label{measure-converge}
\end{equation}
Similarly for $A=[a,b]$ and $\theta_k/2\pi < a \leq b \leq 1$, we obtain
\begin{equation}
 \mu_N(A) \leq \frac{1}{\sin^2[(\theta_k - 2 \pi a)/2]} \frac{2^N(b-a)+1}{2^{2N}}. \label{measure-converge2}
\end{equation}

Let $F_N(f_\mathrm{max})$ be the distribution function of $f(\mathbf{n}_N)$ for a given $N$, i.e.,
\begin{align}
 F_N(f_\mathrm{max}) &= P(f(\mathbf{n}_N) \leq f_\mathrm{max}) \\
 &=  \sum_{f(\mathbf{n}_N) \in [0,f_\mathrm{max}]} p_{N}(f(\mathbf{n}_N) \vert \theta_k)
\end{align}
and $F(f_\mathrm{max})$ be that of the continuous variable $f$, given by
\begin{align}
 F(f_\mathrm{max}) &= P(f(\mathbf{n}_N) \leq f_\mathrm{max}) \\
 &=  \int_0^{f_\mathrm{max}} \delta(\theta_k - 2 \pi f)~df.
\end{align}
For $0 \leq f_\mathrm{max} \leq \theta_k/2\pi$, we have from Ineq.\,\eqref{measure-converge} that
\begin{equation}
  F_N(f_\mathrm{max}) \leq \frac{1}{\sin^2[(\theta_k- 2 \pi f_\mathrm{max})/2]} \frac{2^N f_\mathrm{max}+1}{2^{2N}},
\end{equation} 
therefore,
\begin{equation}
 \lim_{N \rightarrow \infty} F_N(f_\mathrm{max}) = F(f_\mathrm{max}) = 0.
\end{equation}
For $\theta_k/2\pi < f_\mathrm{max} \leq 1$, we see from Ineq.\,\eqref{measure-converge2} that
\begin{equation}
  F_N(f_\mathrm{max}) \geq 1 - \frac{1}{\sin^2[(\theta_k \!-\! 2 \pi f_\mathrm{max})/2]} \frac{2^N (1\!-\! f_\mathrm{max}) \!+\! 1}{2^{2N}},
\end{equation} 
which implies
\begin{equation}
 \lim_{N \rightarrow \infty} F_N(f_\mathrm{max}) = F(f_\mathrm{max}) = 1.
\end{equation}
Therefore, for all points at which $F(f)$ is continuous, $F_N(f)$ converges to $F(f)$, thus the random variable $f(\mathbf{n}_N)$ converges to $f$ in distribution for $N \rightarrow \infty$.

\section{Universal controllization}\label{universal}
\begin{figure}[!]
\vspace{3mm}
	\subfigure{(a)
		\includegraphics[width=0.41\columnwidth]{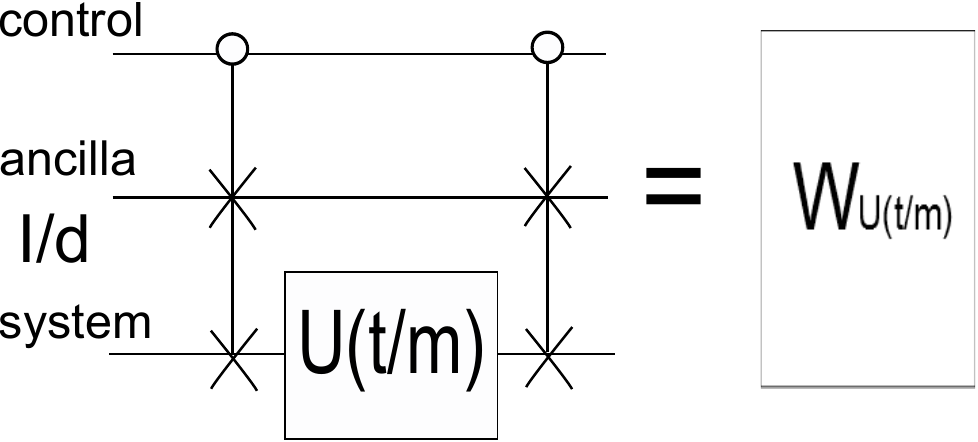}}
\hspace{2mm}
	\subfigure{(b)
		\includegraphics[clip, width=0.41\columnwidth]{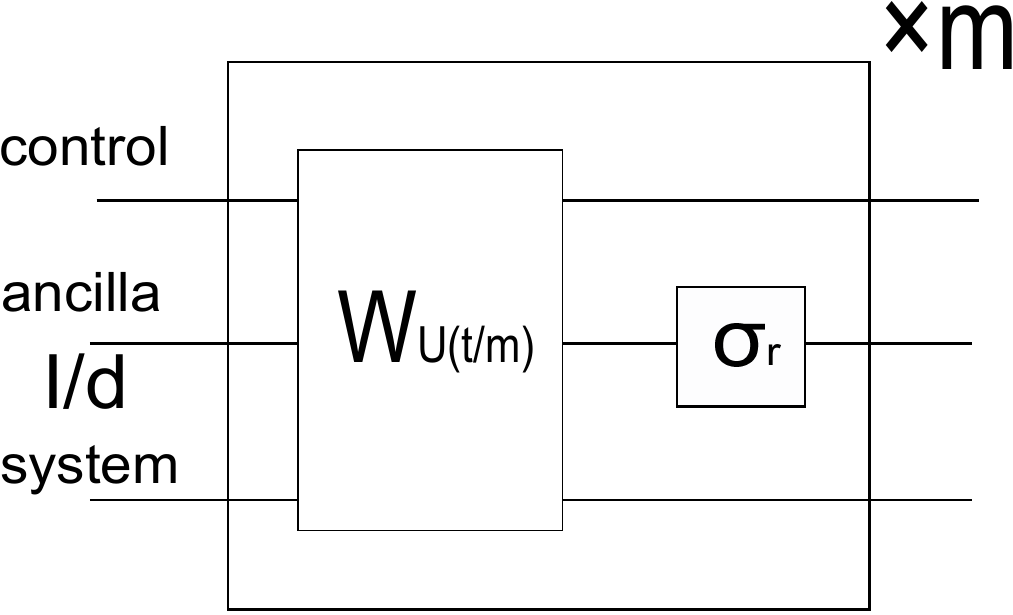}}
 \caption{(a) A quantum circuit representation of the classically conditioned quantum gate $W_{U(t/m)}$.  (b) A quantum circuit representation of the algorithm implementing the universal controllization of $U (t)$. The gate $\sigma_r$ is chosen uniformly randomly for each iteration from the set $S$ defined in Lemma\,\ref{random}. The controlled-swap operation and the random unitary operations are to be performed instantaneously.}
	\label{pseudo}
\end{figure}
This section is related to the fourth section of the main text on {\it QPE and universal controllization'}.
In Sec.~\ref{prel}, we provide a mathematical statement which supports the existence of the refresh operations used in universal controllization.
We derive a description of the superoperator (CPTP map) implemented by the universal controllization in Sec.~\ref{spopcont}, and analyze the error in controllization for a given refresh rate $m$ in Sec.~\ref{accuunicon}.
We also show how to obtain the limit $\lim_{m\rightarrow\infty}m\varphi_{U(t/m)}$ in Sec.~\ref{concentphi}, which appears in universal controllization.
We denote the $k \times k$ identity matrix by $\mathbb{I}_k$.

\subsection{Preliminary}\label{prel}
We present mathematical relations that justify the refreshing operation (3) in the main text.
\begin{lem}\label{random}
Let $\mathcal{H}$ be a $d$-dimensional Hilbert space and $G$ a finite group with a $d \times d$ unitary irreducible representation $\Sigma$.  If a set $S :=\{\sigma_1, \sigma_2, \dots, \sigma_D \}$ of unitaries on $\mathcal{H}$ satisfies
\begin{equation} \label{randomizer}
\Sigma (g) \bigg( \sum_{r=1}^D \sigma_r A \sigma_r^\dagger \bigg) \Sigma (g)^\dagger = \sum_{r=1}^D \sigma_r A \sigma_r^\dagger
\end{equation}
for any operator $A$ on $\mathcal{H}$ and $g \in G$, then
\begin{equation}
 \frac{1}{D}\sum_r \sigma_r A \sigma_r^\dagger =  \frac{\tr{A}}{d} \cdot \mathbb{I}_d
\end{equation}
for any given operator $A$.
\end{lem}

Let us define an operator $\tilde{A} := \sum_{r=1}^D \sigma_r A \sigma_r^\dagger$. 
Equation\,\eqref{randomizer} implies that 
\begin{equation}
 \Sigma (g) \tilde{A}  = \tilde{A} \Sigma (g).
\end{equation}
By Schur's lemma, such an operator $\tilde{A}$ satisfies
\begin{equation}
 \tilde{A} = a \cdot \mathbb{I}_d.
\end{equation} 
Since $\tr{\sigma_r A \sigma_r^\dagger}=\tr{A}$, we have
\begin{equation}
 a = \frac{\tr{\tilde{A}}}{d} = \frac{1}{d \cdot D}\sum_r \tr{\sigma_r A \sigma_r^\dagger}= \frac{\tr{A}}{d},
\end{equation}
which proves Lemma\,\ref{random}.

We also introduce the following corollary of this lemma.
\begin{cor}\label{partialrand}
Let $\mathcal{H}$ be as defined in Lemma\,\ref{random} and $\mathcal{H}^\prime$ be a $d'$-dimensional Hilbert space.  For any operator $M$ on $\mathcal{H}' \otimes \mathcal{H}$ we have that  
\begin{equation}
 \frac{1}{D}  \sum_{r=1}^D \left(\mathbb{I}_{d'}  \otimes \sigma_r \right) M \left( \mathbb{I}_{d'} \otimes \sigma_r\right)^\dagger =  \partialtr{\mathcal{H}}{M} \otimes \frac{\mathbb{I}_d}{d},
\end{equation}
where $\sigma_r$ is taken from $S$ as defined in Lemma\,\ref{random}.
\end{cor}
The proof follows almost immediately from Lemma\,\ref{random}, since any operator $M$ on $\mathcal{H}' \otimes \mathcal{H}$ can be decomposed as
\begin{equation}
 M = \sum_k A'_k \otimes A_k, 
\end{equation}
where $A'_k$ and $A_k$ are operators on $\mathcal{H}'$ and $\mathcal{H}$, {respectively.}

\subsection{Superoperator description}\label{spopcont}
We are now ready to derive the superoperator implemented by the universal controllization. 
Consider a unitary operation $U(t):=\exp(-iHt)$ generated by a Hamiltonian $H$ on $\mathcal{H}_t = \mathbb{C}^d$.
A quantum circuit representation of the algorithm is presented in Fig.\,\ref{pseudo}(b).
It uses one control qubit and a $d$-dimensional ancilla.
The respective Hilbert spaces are denoted by $\mathcal{H}_c$ and $\mathcal{H}_a$.

Let $\rho = \sum_{k,j=0,1}\ket{k}\bra{j} \otimes \rho_{kj}$ on $\mathcal{H}_c \otimes \mathcal{H}_t$ be the initial state of the control-target system.  The initial state of the total system including the ancilla is given by the density matrix
\begin{equation}
\rho_{\rm tot} = \rho \otimes \frac{\mathbb{I}_d}{d}
\end{equation}
on $\mathcal{H}_c \otimes \mathcal{H}_t \otimes \mathcal{H}_a$.
The algorithm first performs the classical conditioned quantum gate
\begin{equation}
 W_{U(t/m)} = C_S \cdot \left( {\mathbb I}_2 \otimes U(t/m) \otimes {\mathbb I}_d \right) \cdot C_S. 
\end{equation}
followed by the refreshing operation $\sigma_r$ on the ancilla.
Here, $C_S$ is the controlled-swap operation, defined in the main text.
Figure\,\ref{pseudo}(b) describes the circuit for pseudo controllization of $U(t/m)$.  Note that we take $W_{U(t/m)}$ as a unitary on $\mathcal{H}_c \otimes \mathcal{H}_t \otimes \mathcal{H}_a$, while the subsystems in the figure are arranged in the order of the control, ancilla, and target, which in the figure is labeled ``system".  These two quantum operations are repeated in the same order for $m$ times.  For each iteration, $\sigma_r$ is chosen uniformly randomly from $S$ defined in Lemma\,\ref{random}.  We see from Corollary\,\ref{partialrand} that the first iteration yields
\begin{multline}
\frac{1}{D}  \sum_r \left( \mathbb{I}_2 \otimes \mathbb{I}_d \otimes \sigma_r \right) W_{U \left(\frac{t}{m}\right)} \rho_{\rm tot}  W_{U\left(\frac{t}{m}\right)}^\dagger \left( \mathbb{I}_2 \otimes \mathbb{I}_d \otimes \sigma_r \right)^\dag \\
= \partialtr{\mathcal{H}_a}{W_{U \left(\frac{t}{m}\right)} \rho_{\rm tot}  W_{U\left(\frac{t}{m}\right)}^\dagger} \otimes \frac{\mathbb{I}_d}{d} = \Gamma_{U\left(\frac{t}{m}\right)}[ \rho] \otimes \frac{\mathbb{I}_d}{d} ,
\end{multline}
where the summation over $r$ is to reflect that $\sigma_r$ is chosen uniformly randomly. 
Thus, $m$ iterations achieve
\begin{equation}
 \rho \otimes \frac{\mathbb{I}_d}{d} \rightarrow \Gamma^m_{U\left(\frac{t}{m}\right)}[ \rho] \otimes \frac{\mathbb{I}_d}{d}.
\end{equation}

Simple algebra will show that
\begin{align}
&W_{U \left(\frac{t}{m}\right)} \rho_{\rm tot}  W_{U\left(\frac{t}{m}\right)}^\dagger \nonumber\\
& = \frac{1}{d}  \sum_{k,j} \ket{k}\bra{j} \otimes  U \left (\frac{kt}{m} \right ) \rho_{kj} U^\dagger \left (\frac{jt}{m} \right ) \otimes U \left (\frac{(j-k)t}{m} \right ).
\end{align}
Therefore, the first iteration can be seen as transformation
\begin{align}
 \rho_{00} &\rightarrow \rho_{00},\\
 \rho_{01} &\rightarrow \rho_{01}\Big(\gamma_{U(t/m)}U^\dag(t/m)\Big),\\
 \rho_{10} &\rightarrow \Big(\gamma^*_{U(t/m)}U(t/m)\Big)\rho_{10},\\
 \rho_{11} &\rightarrow U(t/m) \rho_{11} U^\dag(t/m),
\end{align}
where
\begin{equation}
 \gamma_{U(\tau)} := \tr{U(\tau)}/d.
\end{equation}
We have thus
\begin{multline}
 \Gamma_{U\left(\frac{t}{m}\right)}^m[\rho] =  C_{U(t)} \rho C_{U(t)}^\dag + \\
 \left[ \ket{0}\bra{1} \otimes \rho_{01} \left( \gamma^m_{U(t/m)} -1 \right) U^\dag + c.c.\right],
\end{multline}
which is the superoperator implemented by the universal controllization.

\subsection{Accuracy of the universal controllization}\label{accuunicon}
The previous subsection shows that the universal controllization is a map from a quantum gate $U(t)$ to the superoperator $\Gamma_{U\left(t/m\right)}^m$.
The ideal universal controllization would be a map from $U(t)$ to the superoperator \begin{equation}
 \mathcal{C}_{U(t)}[\rho] := C_{U(t)} \rho C_{U(t)}^\dag.
\end{equation}
Let us evaluate the accuracy of the universal controllization for a given $m$ as a distance between the maps $\Gamma_{U(t/m)}^m$ and $\mathcal{C}_{U(t)}$.

\begin{theo}\label{distance}
For any $m \in \mathbb{N}$ and a unitary operator $U(t)=\exp(-iHt)$ generated by a Hermitian operator $H$ on $\mathbb{C}^d$ and $t \in \mathbb{R}$, we have
\begin{equation}
 \norm{\Gamma_{U(t/m)}^m - \mathcal{C}_{U^{[m]}(t)} }_\diamond = 1 - a_{U(t/m)}^m,
\end{equation}
where
\begin{equation}
 a_{U(\tau)} := \abs{\frac{\tr{U(\tau)}}{d}}, \blank U^{[m]}(t) := e^{-im\varphi_{U(t/m)}}U(t),
\end{equation}
for $\varphi_{U(\tau)}$ defined by
\begin{equation} \label{phase}
 e^{i\varphi_{U(\tau)}} := \frac{\gamma_{U(\tau)}}{a_{U(\tau)}}.
\end{equation} 
\end{theo}

The \textit{diamond norm} $\norm{\cdot}_\diamond$\,\cite{Diamond} in this theorem is a norm for superoperator, which takes into account when the {superoperators} is extendend to act on a part of a larger Hilbert space than for which it is originally defined.
It is often used to evaluate the difference between two CPTP maps in the context of quantum information.

A superoperator ${\mathcal S}$ on a Hilbert space ${\mathcal H}$ acting on an extended system
${\mathcal H}\otimes{\mathcal H}^\prime$ satisfies
\begin{equation}
 \norm{\mathcal S}_{\rm op} < \norm{\left( {\mathcal S} \otimes {\mathrm{id}}_{{\mathcal H}^\prime} \right)}_{\rm op},
\end{equation}
where the operator norm $\norm{\mathcal S}_{\rm op}$ is the maximum of the trace norm  
of ${\mathcal S} [A]$ for an operator $A$ under the condition $\norm{A}_{\rm tr}=1$ and
${\mathrm{id}}_{{\mathcal H}^\prime}$ denotes the identity superoperator on ${\mathcal H}^\prime$.
The trace norm is defined as $\norm{A}_{\rm tr} = \tr{AA^\dagger}{}$.
Since $\norm{{\mathcal S}\otimes {{\mathrm{id}}}_{{\mathcal H}^\prime}}_{\rm op} \leq \norm{{\mathcal S}\otimes {\mathrm{id}}_{\mathcal H}}_{\rm op}$ holds for any Hilbert space ${\mathcal H}^\prime$, it is enough to consider $\norm{{\mathcal S}\otimes {\mathrm{id}}_{\mathcal H}}_{\rm op}$ to bound $\norm{\mathcal{S}\otimes {\rm id}_{\mathcal{H}'}}$ for any $\mathcal{H}'$.
The diamond norm $\norm{\cdot}_\diamond$ of a superoperator ${\mathcal S}$ 
on the Hilbert space ${\mathcal H}$ is defined as
\begin{equation}
 \norm{{\mathcal S}}_{\diamond} := \norm{{\mathcal S}\otimes {\mathrm{id}}_{\mathcal H}}_{\rm op}.
\end{equation}

The following lemma is convenient for calculating the diamond norm.
\begin{lem}\label{dialem}
Any Hermitian preserving superoperator $\Lambda$ on the Hilbert space $\mathcal{H}$ satisfies
\begin{equation}
 \norm{\Lambda}_{\diamond} = \max_{P \in {\mathcal P}_1} \norm{\left( \Lambda \otimes {\mathrm{id}}_{\mathcal H}  \right) P}_{\rm tr},
\end{equation}
where ${\mathcal P}_1$ is a set of rank-1 projectors on ${\mathcal H}\otimes{\mathcal H}$.
\end{lem}
\noindent See Ref.\,\cite{Diamond} for a proof.

Let us prove Theorem\,\ref{distance}. 
To calculate the diamond norm, we search for rank-1 projectors on $(\mathcal{H}_c \otimes \mathcal{H}_t)^{\otimes 2} := (\mathcal{H}_c \otimes \mathcal{H}_t) \otimes (\mathcal{H}_c \otimes \mathcal{H}_t)$ that gives the largest trace norm after $\Gamma^m_{U(t/m)}-\mathcal{C}_{U^{[m]}(t)}$ is applied.
Any rank-1 projector on $(\mathcal{H}_c \otimes \mathcal{H}_t)^{\otimes 2}$ is given by $\ket{\Psi}\bra{\Psi}$ for some vector $\ket{\Psi}$ in $(\mathcal{H}_c \otimes \mathcal{H}_t)^{\otimes 2}$.

All vectors in $(\mathcal{H}_c \otimes \mathcal{H}_t)^{\otimes 2}$ can be represented as $\ket{\Psi} = \alpha \ket{0}\ket{\psi} + \beta \ket{1}\ket{\phi}$, where $\{\ket{0}, \ket{1} \}$ is the computational basis of the first control qubit system $\mathcal{H}_c$, $\ket{\psi}$ and $\ket{\phi}$ are normalized vectors in $\mathcal{H}_t \otimes \mathcal{H}_c \otimes \mathcal{H}_t$, and $\alpha, \beta$ satisfy $|\alpha|^2+ |\beta|^2 =1$.  As a block matrix, the projector is represented by 
\begin{equation}
 \ket{\Psi}\bra{\Psi} = 
	\begin{pmatrix}
	\abs{\alpha}^2 \ket{\psi}\bra{\psi} && \alpha \beta^\ast \ket{\psi}\bra{\phi} \\
	\alpha^\ast \beta \ket{\phi}\bra{\psi} && \abs{\beta}^2 \ket{\phi}\bra{\phi} 
	\end{pmatrix}.
\end{equation}
The upper left block corresponds to the $\ket{0}\bra{0}$ element of the first system.
The upper right block is  the $\ket{0}\bra{1}$ element, and the other blocks are defined similarly.
The projector $\ket{\Psi}\bra{\Psi}$ is transformed by the maps ${\mathcal C}_{U^{[m]}(t)}$ and $\Gamma_{U({t}/{m} )}^{m}$ as
\begin{align}
 &\left({\mathcal C}_{U^{[m]}(t)} \otimes {\mathrm{id}}_{{\mathbb C}^2 \otimes {\mathbb C}^d} \right) [\ket{\Psi}\bra{\Psi}] \nonumber\\
 &\quad=
	\begin{pmatrix}
	\abs{\alpha}^2 \ket{\psi}\bra{\psi} &&  \alpha \beta^\ast \ket{\psi}\bra{\phi}U^{[m]}(t)^\dagger \\
	\alpha^\ast \beta U^{[m]}(t)\ket{\phi}\bra{\psi} && \abs{\beta}^2 U^{[m]}(t)\ket{\phi}\bra{\phi} U^{[m]\dagger}(t)
	\end{pmatrix}
\end{align}
and
\begin{align}
 &\left({\Gamma_{U\left(\frac{t}{m} \right)}^{m}} \otimes {\mathrm{id}}_{{\mathbb C}^2 \otimes {\mathbb C}^d } \right)[\ket{\Psi}\bra{\Psi}] \nonumber\\
&= 
	\begin{pmatrix}
	\abs{\alpha}^2 \ket{\psi}\bra{\psi} &&  \gamma_{U\left(\frac{t}{m}\right)}^m \alpha \beta^\ast \ket{\psi}\bra{\phi}U(t)^\dagger \\
	 \gamma_{U\left(\frac{t}{m}\right)}^{\ast m} \alpha^\ast \beta U(t)\ket{\phi}\bra{\psi} && \abs{\beta}^2 U(t)\ket{\phi}\bra{\phi} U(t)^\dagger
	\end{pmatrix}.
\end{align}
Note that 
\begin{align}
 \gamma^m_{U\left(\frac{t}{m} \right)} &= \exp\left(im \varphi_{U\left(\frac{t}{m} \right)}\right) a_{U\left(\frac{t}{m} \right)}^{m},\\
 U^{[m]}(t) &= \exp\left({-im\varphi_{U\left(\frac{t}{m} \right)}} \right)U(t).
\end{align} 
A direct calculation will show that
\begin{equation}
\norm{
	\begin{pmatrix}
	0 && \alpha \beta^\ast \ket{\psi}\bra{\phi}U^{[m](t)\dagger} \\
	\alpha^\ast \beta U^{[m]}(t)\ket{\phi}\bra{\psi} && 0
	\end{pmatrix}
	}_{\rm tr}
= 2 \abs{\alpha \beta}.
\end{equation}
Therefore, the norm of interest is 
\begin{multline}
 \norm{{\mathcal C}_{U^{[m]}(t)} - \Gamma_{U\left(\frac{t}{m} \right)}^{m}}_\diamond \\
 = 2\left(1-a_{U\left(\frac{t}{m} \right)}^m\right) \max_{\alpha,\beta }  \abs{\alpha \beta} = 1- a_{U\left(\frac{t}{m} \right)}^m,\label{errorcont}
\end{multline}
where we have used the normalization condition for $\alpha$ and $\beta$ to obtain the last equality.
This proves Theorem\,\ref{distance}.

The distance between ${\mathcal C}_{U^{[m]}(t)}$ and $\Gamma_{U({t}/{m} )}^{m}$ approaches $0$ as $m$ increases.  %An intuitive argument is as follows.  
{Here is an intuitive argument:}  First,
\begin{align}
 &a_{U(t/m)} = \sqrt{\frac{1}{d} \tr{U(t/m)} \cdot \tr{U(t/m)}^\ast} \nonumber\\
 &\qquad = \sqrt{\frac{1}{d^2} \sum_k e^{-iE_kt/m} \cdot \sum_l e^{iE_lt/m}} \nonumber\\
 &\qquad = \sqrt{\frac{1}{d^2} \sum_{k,l} e^{i(E_k-E_l)t/m}} \nonumber\\
 &\qquad = \sqrt{\frac{2}{d^2} \left(\sum_{k> l}\cos \big[(E_k - E_l)t/m\big] \right) + \frac{1}{d}}.
\end{align}
We invoke the Taylor expansion of $\cos(\alpha x)$ to the second {order in $x$.}%power of $x$,
\begin{equation}
 \cos(ax) = 1 - \frac{\alpha^2}{2} x^2 + O(x^4),
\end{equation}
and let $\alpha_{kl} =  (E_k - E_l)t$ and $x = 1/m$.  Under these notations,
\begin{align}
 &a_{U(t/m)} = \sqrt{1 - \frac{1}{d^2} \left( \sum_{k>l} \alpha^2_{kl} \bigg(\frac{1}{m} \bigg)^2\right)  +  O(m^{-4})} \nonumber\\
& \qquad =1-C\bigg(\frac{1}{m} \bigg)^2+O(m^{-4}),
\end{align}
where the last equality is derived using the Taylor expansion of $\sqrt{1+x}$ to the second order of $x$ with $C = \left( \sum_{k>l} \alpha^2_{kl} \right)/2d^2$.  Finally, we Taylor expand  $(1+x)^m$ to the second {order in} $x$ and set $x = C/m^2$ to obtain 
\begin{equation}
 a^m_{U(t/m)} = 1-m \cdot C\bigg(\frac{1}{m} \bigg)^2+O(m^{-3}),
\end{equation}
which converges to $1$ as $m \rightarrow \infty$.  Thus the distance \eqref{errorcont} converges to $0$.
{This property is mathematically analogous to the quantum {Zeno} effect\,\cite{Zeno}.}

\subsection{Convergence of the phase factor}\label{concentphi}
We stated at the end of the fourth section in the main text (``Universal controllization") that the phase shift $m \varphi_{U(t/m)}$ induced by the universal controllization satisfies 
\begin{equation}
 \lim_{m \rightarrow \infty} e^{i m \varphi_{U(t/m)}} = e^{-i\tr{H}t/d}.
\end{equation}
A proof is as follows.
Since $a_{U(t/m)} = 1 + O(1/m^2)$, the coherence factor can be sorted by the order of $m$ as 
\begin{equation} 
 \Big(e^{i\varphi(t/m)}\Big)^m = \left(1 - i \frac{\tr{H}{}}{d} \frac{t}{m} + O\left(\frac{1}{m^2}\right) \right)^m.
\end{equation}
Hence, we can conclude that 
\begin{eqnarray} 
 e^{im\varphi(t/m)} &=& \left(1 - i \frac{\tr{H}{}}{d}\frac{t}{m}  \right)^m + O\left(\frac{1}{m}\right)  \\
	&=& e^{- i \tr{H}t/d} + O\left(\frac{1}{m}\right).
\end{eqnarray}

{\section{Algorithm for directly evaluating the accuracy of controllization}}
In the main article, universal controllization is introduced as a subroutine for PME. However, applications of universal controllization is not limited to PME, since many algorithms and protocols in quantum information utilize controlled-unitary operations.  In this section, we introduce a quantum algorithm that directly evaluates the accuracy of universal controllization given by Eq.~(\ref{errorcont}).

\begin{figure}[t]
 \begin{center}
\vspace{8mm}
  \includegraphics[clip, width=0.8\columnwidth]{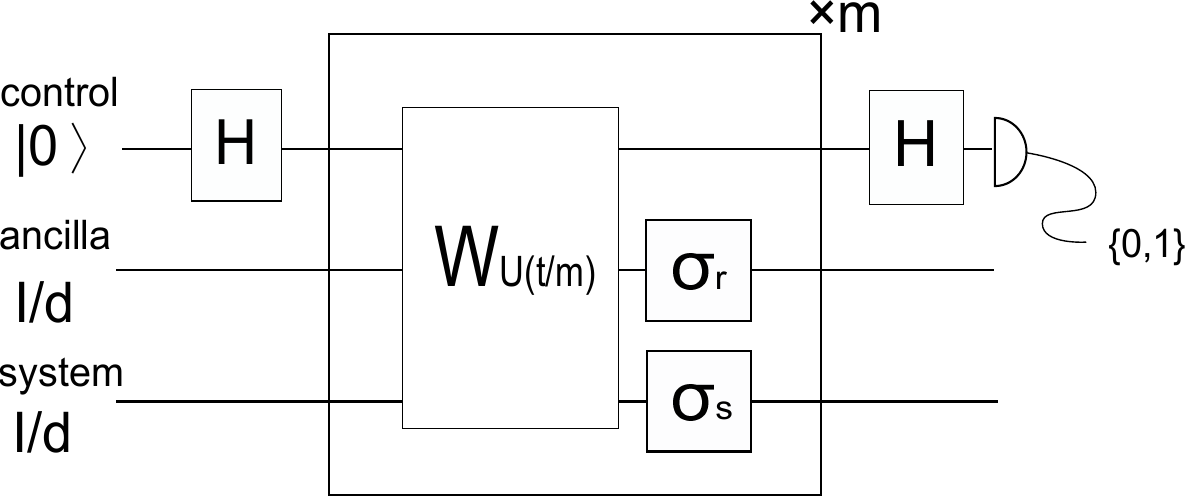}
 \end{center}
 \caption{A quantum circuit representation of the algorithm that estimates the controllization error of the universal controllization due to finite refresh rate $m$.}\label{rADQC1}
\end{figure}

Figure\,\ref{rADQC1} gives a quantum circuit representation of the algorithm.
The total system consists of three subsystems, namely, the control ($\mathcal{H}_c$), target ($\mathcal{H}_t$), and ancilla ($\mathcal{H}_a$), with dimension $2$, $d$, and $d$, respectively. 
We shall use $\mathbb{I}_k$ to denote the $k \times k$ identity matrix.
The system is prepared in the state
\begin{equation} \label{afterHad}
 \ket{0}\bra{0} \otimes \frac{\mathbb{I}_d}{d} \otimes \frac{\mathbb{I}_d}{d}
\end{equation}
on $\mathcal{H}_c \otimes \mathcal{H}_t \otimes \mathcal{H}_a$.
We denote the Pauli X, Y, and Z matrix as $\sigma_x$, $\sigma_y$, and $\sigma_z$, respectively, whose matrix representation is
\begin{align}
 \sigma_x &= \ket{0}\bra{1} + \ket{1}\bra{0}\\
 \sigma_y &= -i\ket{0}\bra{1} + i\ket{1}\bra{0}\\
 \sigma_z &= \ket{0}\bra{0} - \ket{1}\bra{1}.
\end{align} 

Our goal is to convert the state of the control qubit to 
\begin{equation} \label{goalstate}
 \rho_m = \frac{1}{2} \left( \mathbb{I}_2 + a_{U(t/m)}^{2m} \sigma_z \right)
\end{equation}
and obtain the expectation value for $\sigma_z$, i.e.,
\begin{equation} \label{goalvalue}
\mathrm{Tr}[ {\rho}_m \sigma_z] = a_{U(t/m)}^{2m}.
\end{equation}
First, we apply the Hadamard gate $\mathrm{H}$ on the control qubit.  This transforms the total state to
\begin{equation}
 \frac{\mathbb{I}_2 + \sigma_x}{2} \otimes \frac{\mathbb{I}_d}{d} \otimes \frac{\mathbb{I}_d}{d}.
\end{equation}

Consider a set of unitary operations $S=\{ \sigma_r \vert 1 \leq r \leq D\}$ on $\mathbb{C}^d$ as defined in Lemma\,\ref{random} and randomly choose a unitary operation \begin{equation}
 \sigma_{rs} : = \sigma_r\otimes \sigma_s.
\end{equation}
We perform the unitary operation $W_{U(t/m)}$ on the state Eq.\,\eqref{afterHad}, which is followed by $\mathbb{I}_d \otimes \sigma_{rs}$.  Each of these operations is applied $m$ times in total, while $\sigma_{rs}$ is chosen at random for each repetition. 

Note that 
\begin{multline}
 W_{U} \left(\sigma_x \otimes \mathbb{I}_d \otimes \mathbb{I}_d \right)W_{U}^\dagger \\
= \sigma_x \otimes \left(U \otimes U^\dagger + U^\dagger \otimes U\right)  \\
- i\sigma_y \otimes \left(U \otimes U^\dagger - U^\dagger \otimes U\right).
\end{multline}
It is also easy to see that
\begin{align}
&\frac{1}{D^2}\sum_{r,s} \left(\mathbb{I}_2 \otimes \sigma_{rs} \right) W_{U} \left(\sigma_x \otimes \mathbb{I}_d \otimes \mathbb{I}_d \right)W_{U}^\dagger \left(\mathbb{I}_2 \otimes \sigma_{rs} \right)^\dagger  \nonumber \\ 
&\qquad = \frac{\abs{\tr{U}}^2}{d^2} \cdot \sigma_x \otimes \mathbb{I}_d \otimes \mathbb{I}_d \nonumber \\
&\qquad = a^2_U \cdot \sigma_x \otimes \mathbb{I}_d \otimes \mathbb{I}_d.  
\end{align}

The first iteration of $W_{U(t/m)}$ and $\mathbb{I} \otimes \sigma_{rs}$ converts the state Eq.\,\eqref{afterHad} to 
\begin{equation}
 \frac{\mathbb{I}_2 + a_{U(t/m)}^2 \sigma_x}{2} \otimes \frac{\mathbb{I}_d}{d} \otimes \frac{\mathbb{I}_d}{d}.
\end{equation}
The next iteration changes $a_{U(t/m)}^2 \sigma_x$ to $a_{U(t/m)}^4 \sigma_x$.
Therefore, $m$ iterations create the state
\begin{equation}
 \frac{\mathbb{I}_2 + a_{U(t/m)}^{2m} \sigma_x}{2} \otimes \frac{\mathbb{I}_d}{d} \otimes \frac{\mathbb{I}_d}{d}.
\end{equation}
We have the desired state \eqref{goalstate} with another Hadamard gate on the control qubit.

The expectation value \eqref{goalvalue} is calculated from the equation
\begin{equation}
 \mathrm{Tr}[ \rho_m \sigma_z] = \bra{0}\rho_m\ket{0} - \bra{1}\rho_m\ket{1}.
\end{equation}
Each term in the right hand side is the probability of obtaining the outcome 0 and 1, respectively, from the measurement on the control qubit in the computational basis.
Thus, our algorithm calculates the distance \eqref{errorcont}.

\section{Approximated QPE with universal controllization}\label{apxqpe}

This section provides a supplemental material for the fifth section in the main article on {\it  Implementation accuracy and time cost}.  In this section, we follow the calculations evaluating the error of approximated QPE implemented by universal controllization.  
The error is defined as the deviation of the probability distribution of QPE from the ideal case.
In Sec.~\ref{probout}, we first derive the probability distribution of the approximated QPE.  
Then we evaluate the deviation of the probability distribution from the ideal distribution in Sec.~\ref{AAqpe}.  

\vspace{0.6cm}

\subsection{Probability of outcomes}\label{probout}
Let us calculate the probability distribution $p^{[m]}_N(f(\mathbf{n}_N) \vert E_k)$ of {QPE} with universal controllization.
The initial state of the control ($\mathcal{H}_\mathrm{cont}$) and target system is given by
\begin{equation}
 \ket{0 \dots 0} \bra{0 \dots 0} \otimes \ket{E_k} \bra{E_k},
\end{equation}
on $\mathcal{H}_\mathrm{cont} \otimes \mathcal{H}_t$.  We note that $\dim \mathcal{H}_\mathrm{cont} = 2^N$.

At the first step of the algorithm, the Hadamard gate is applied to each control qubit system. 
The state after this operation is given by
\begin{equation}
 \frac{1}{2^N}\sum_{\begin{subarray}aa_1,a_2,\dots a_N \\ b_1, b_2 ,\dots, b_N\end{subarray}} \ket{a_1a_2 \dots a_N}\bra{b_1 b_2 \dots b_N} 
	\otimes \ket{E_k}\bra{E_k}
\end{equation}
where $a_l, b_l \in \{0,1 \}$. 

At the second step, the universal controllization map $\Gamma_{U^{2^{l-1}}(t/m)}^{m}$ is applied on the pair of the $l$-th control qubit and the target system, for all $1 \leq l \leq N$.
These operations transform the state to 
\begin{multline}
 \frac{1}{2^N}\sum_{\begin{subarray} aa_1,\dots a_N \\ b_1, \dots, b_N\end{subarray}} 
	\prod_{l=1}^N a_{U\left(\frac{t}{m}\right)}^{m2^{l-1} \abs{a_l - b_l}} \\ 
	\times \exp\left( i2^{l-1} (a_l - b_l) \left( -E_k t + m \varphi_{U\left(\frac{t}{m}\right)} \right)\right) \\
	\times \ket{a_1, a_2 \dots a_N}\bra{b_1 b_2 \dots b_N} 
	\otimes \ket{E_k}\bra{E_k}.\label{midprot}
\end{multline}
Using the periodicity of the phase function and $\theta'_k =  -E_k t + m \varphi_{U\left(\frac{t}{m}\right)} \pmod{2\pi}$, the above equation simplifies to
\begin{multline}
 \frac{1}{2^N}\sum_{\begin{subarray} aa_1,\dots a_N \\ b_1, \dots, b_N\end{subarray}} 
	\prod_{l=1}^N a_{U\left(\frac{t}{m}\right)}^{m2^{l-1} \abs{a_l - b_l}} \exp\left( i2^{l-1} (a_l - b_l) \theta'_k \right)\\
 \times \ket{a_1, a_2 \dots a_N}\bra{b_1 b_2 \dots b_N} 
	\otimes \ket{E_k}\bra{E_k}.\label{midprot2}
\end{multline}
Unlike Eq.\,\eqref{aftersecond}, the coherence factor $a_{U\left(\frac{t}{m}\right)}$ appears in Eq.\,\eqref{midprot2}.

In the final step, the inverse quantum Fourier transformation is applied and the control qubits are measured in the computational basis.
This is equivalent to performing a projective measurement in the Fourier basis \eqref{fourierbasis}.
The probability $p^{[m]}_N(f(\mathbf{n}_N)|E_k)$ of obtaining $f(\mathbf{n}_N)$ by the measurement 
$\left \{\ket{f(\mathbf{n}_N)}  \vert 0 \leq \mathbf{n}_N < 2^{N} \right \}$
on a density operator $\rho$ is given by $\bra{f(\mathbf{n}_N)} \rho \ket{f(\mathbf{n}_N)}$.
Therefore, according to Eq.~(\ref{midprot2}) the probabilty distribution of {QPE} with universal controllization satisfies 
\begin{align}
  &p^{[m]}_N(f(\mathbf{n}_N) \vert E_k) \nonumber\\
 &= \frac{1}{2^{2N}} \sum_{\begin{subarray} aa_1,\dots a_N \\ b_1,\dots, b_N\end{subarray}} \prod_{l=1}^N a_{U\left(\frac{t}{m}\right)}^{m2^{l-1}\abs{a_l - b_l}} \\
&\qquad\qquad \times \exp\left({i2^{l-1} (a_l - b_l) \left( \theta'_k -2\pi f(\mathbf{n}_N) \right) }\right) \nonumber \\
 &=   \frac{1}{2^{2N}}\prod_{l=1}^N  \sum_{a,b = 0,1} a_{U\left(\frac{t}{m}\right)}^{m2^{l-1}\abs{a - b}} \nonumber \\
&\qquad\qquad \times \exp\left({i2^{l-1} (a - b) \left( \theta'_k - 2\pi f(\mathbf{n}_N) \right) } \right) \\
 &= \frac{1}{2^N} \prod_{l=1}^N  \left[ 1 + a_{U\left(\frac{t}{m}\right)}^{m2^{l-1}} \cos \big[ 2^{l-1} \left( \theta_k^\prime - 2\pi f(\mathbf{n}_N) \right) \big]\right].
\end{align}

\subsection{Accuracy of the approximated {QPE}}\label{AAqpe}
{The error of approximated QPE is evaluated as $\abs{p_N^{[m]} (f(\mathbf{n}_N) | E_k) - p_N (f(\mathbf{n}_N)\vert \theta_k')} $.
To calculate this error}, we define the following quantity $\delta_l$ representing the effect of the phase damping noise,
\begin{equation}
 \delta_l = 1- a_{U(t/m)}^{m2^{l-1}}.
\end{equation}
The probability distribution $p^{[m]}(f(\mathbf{n}_N)\vert E_k)$ can be expressed as
\begin{equation}
 p_N^{[m]} (f(\mathbf{n}_N) | E_k)  = \prod_{l=1}^N  ( \mu_l - \delta_l \nu_l ), \label{deltadef2}
\end{equation}
where
\begin{align}
 \mu_l &= \frac{1 + \cos {2^{l-1} \left( \theta_k^\prime - 2\pi f(\mathbf{n}_N) \right) }}{2}\\
 \nu_l &= \frac{\cos 2^{l-1} \left( \theta_k^\prime - 2\pi f(\mathbf{n}_N) \right) }{2}.
\end{align}

We define a set $K(l)$ of subsequences of $\{1,2,\dots,N\}$ of length $l$, such that
\begin{multline}
 K(l) : = \Big\{ \{m_1,m_2,\dots,m_l \} \big\vert \forall i, 1 \leq m_i <m_{i+1} \leq N \Big\}.
\end{multline}
The decomposition of Eq.~(\ref{deltadef2}) is given as
\begin{multline}
 p_N^{[m]} (f(\mathbf{n}_N) | E_k)  \\
=  \prod_{l=1}^N  \mu_l + \sum_{l=1}^N\sum_{\mathcal{K} \in K(l)} \prod_{i \notin \mathcal{K}} \mu_i \prod_{j \in \mathcal{K}} \delta_j \nu_j . 
\end{multline}
Note that the first term is $p_N(f(\mathbf{n}_N)\vert \theta_k')$. Then
\begin{multline}
 \abs{p_N^{[m]} (f(\mathbf{n}_N) | E_k) - p_N (f(\mathbf{n}_N)\vert \theta_k') }  \\
=  \abs{\sum_{l=1}^N\sum_{\mathcal{K} \in K(l)} \prod_{i \notin \mathcal{K}} \mu_i \prod_{j \in \mathcal{K}} \delta_j \nu_j }.
\end{multline}
By the triangular inequality, we have
\begin{multline}
 \abs{\sum_{l=1}^N\sum_{\mathcal{K} \in K(l)} \prod_{i \notin \mathcal{K}} \mu_i \prod_{j \in \mathcal{K}} \delta_j \nu_j }\\
 \leq \sum_{l=1}^N\sum_{\mathcal{K} \in K(l)} \prod_{i \notin \mathcal{K}} \abs{\mu_i} \prod_{j \in \mathcal{K}} \delta_j \abs{\nu_j}. 
\end{multline}
Since $\abs{\mu_l}\leq 1$ and $\abs{\nu_l}\leq 1/2$, we arrive at
\begin{equation}
 \abs{p_N^{[m]} (f(\mathbf{n}_N) | E_k) - p_N (f(\mathbf{n}_N)\vert \theta_k') }  
\leq  \sum_{l=1}^N \sum_{\mathcal{K} \in K(l)} \frac{\prod_{j \in \mathcal{K}} \delta_j}{2^l}. 
\end{equation}

Let us assume the condition
\begin{equation}
 \delta_N = 1 -a_{U(t/m)}^{m 2^{N-1}} \leq \delta \label{conda}, 
\end{equation} 
is satisfied for a fixed, then $\delta_l \leq \delta $ is satisfied for $ 1\leq l \leq N $ since $a_{U(t/m)}^m \leq 1$. 
Therefore
\begin{multline}
 \abs{p_N^{[m]} (f(\mathbf{n}_N) | E_k) - p_N (f(\mathbf{n}_N)\vert \theta_k') }  \\
\leq  \sum_{l=1}^N \sum_{\mathcal{K} \in K(l)} \frac{\delta^l}{2^l} 
=  \sum_{l=1}^N \frac{N!}{l!(N-l)!} \cdot \frac{\delta^l}{2^l} \\
=  \left(1 + \frac{\delta}{2}\right)^N-1. \label{iooptemae}
\end{multline}

Here we introduce the following lemma.
\begin{lem}\label{expoasy}
 For any $0\leq a \leq 1$ and $0 \leq r \leq r' $, 
\begin{eqnarray}
 \left( 1 - \frac{a}{r}\right)^{r} \leq \left( 1 - \frac{a}{r'}\right)^{r'}, \\
 \left( 1 + \frac{a}{r}\right)^{r} \geq \left( 1 + \frac{a}{r'}\right)^{r'}.
\end{eqnarray} 
\end{lem}
This lemma can be easily checked by the following differential relation,
\begin{eqnarray}
 \frac{d}{dr}\left( 1 - \frac{a}{r}\right)^r = \frac{a}{r}\left( 1 - \frac{a}{r}\right)^{r-1} \geq 0, \\
 \frac{d}{dr}\left( 1 + \frac{a}{r}\right)^r = -\frac{a}{r}\left( 1 + \frac{a}{r}\right)^{r-1} \leq 0.
\end{eqnarray}
We obtain
\begin{equation}
 \abs{p_N^{[m]} (f(\mathbf{n}_N) | E_k) - p_N (f(\mathbf{n}_N)\vert \theta_k') } \leq \frac{N\delta}{2},\label{b}
\end{equation}
from Eq.\,\eqref{iooptemae} by using Lemma \ref{expoasy} and setting $a=N\delta/2$ and $r'=N,r=1$.

Next, we derive sufficiently large $m$ to bound the right-hand side of Eq.~(\ref{b}) by $\varepsilon$.
First, we transform $a_{U(t)}$ as
\begin{eqnarray}
 a_{U(t)}^2 &=& \frac{1}{d} \tr{U(t)} \cdot \tr{U(t)}^\ast \\
 &=& \frac{1}{d^2} \sum_k e^{-iE_kt} \cdot \sum_l e^{iE_lt} \\
 &=& \frac{1}{d^2} \sum_{k,l} e^{i(E_k-E_l)t} \\
 &=& \frac{2}{d^2} \sum_{k> l}\cos (E_k - E_l)t + \frac{1}{d}. \label{cos} 
\end{eqnarray}
Note that $\cos x \geq 1 -x^2/2 $.  By using the maximum difference $\Delta_\mathrm{max} = \max_{k,l} |E_k - E_l|$ in energy eigenvalues, we have 
\begin{eqnarray}
 a_{U(t)}^2 &\geq& 1-  \sum_{k < l} \left( \frac{E_k -E_l}{d} \right)^2t^2 \\
 &\geq& 1 - \frac{d(d-1)}{2d^2}\Delta_{\rm max}^2t^2.
\end{eqnarray}
The inequality above gives the following inequality,
\begin{equation}
 a_{U(t/m)}^{m2^{N-1}} \geq \left(1 - \frac{\Delta_{\rm max}^2t^2}{2m^2} \right)^{m2^{N-1}}.
\end{equation}
From Lemma\,\ref{expoasy},
\begin{equation}
 a_{U(t/m)}^{m2^{k-1}} \geq \left(1 - \frac{\Delta_{\rm max}^2t^2}{2m^2} \right)^{m2^{N-1}} \geq 1 - 2^k\cdot \frac{\Delta_{\rm max}^2t^2}{4m},
\end{equation}
thus
\begin{equation}
 \delta_N = 1 - a_{U(t/m)}^{m2^{N-1}} \leq 2^N \cdot \frac{\Delta_{\rm max}^2t^2}{4m}. \label{chokuzen}
\end{equation}
By Eqs.~(\ref{conda}), (\ref{b}) and (\ref{chokuzen}), we have
\begin{equation}
  \abs{p_N^{[m]}(f(\mathbf{n}_N)\vert E_k ) -p_N(f(\mathbf{n}_N)\vert \theta_k^\prime)} \leq 2^{N-3} \cdot \frac{\Delta_{\rm max}^2 t^2N}{m}.
\end{equation}
Thus for any $\varepsilon > 0$, if 
\begin{equation}
 m \geq \frac{\Delta_{\rm max} ^2 t^2 N 2^{N-3}}{\varepsilon},
\end{equation}
then
\begin{equation}
 \abs{p_N^{[m]}(f(\mathbf{n}_N)\vert E_k ) -p_N(f(\mathbf{n}_N)\vert \theta_k^\prime)} \leq \varepsilon.
\end{equation}

\section{Estimating the energy scale}\label{ES}
We claimed that we can find an upper bound of $\Delta_\mathrm{max}$ by estimating $a_{U(t)}$ for some prior distribution of the system Hamiltonian $H$ in the sixth section in the main article on {\it Quantum algorithmic estimation of the energy scale}. 
In {subsection \ref{protocolES}}, we {present} the protocol {for estimating of the energy scale}.  This protocol employs a subroutine algorithm evaluating $a_U$ introduced in Sec.~\ref{esta}.  
Evaluation of the failure probability of the protocol is shown in Sec.~\ref{probinc} together with mathematical formulas used in evaluation.

\vspace{0.5cm}
\subsection{Protocol for estimating of the energy scale} \label{protocolES}

Suppose we have a $d$-dimensional quantum system for  $d\geq3$.
A precise description first involves regarding $H$ as a random variable.
We denote the minimum and maximum energy eigenvalue of $H$ by $E_\mathrm{min}$ and  $E_\mathrm{max}$, respectively, which are both a random variable themselves.
Let $F(E_\mathrm{min}\!=\!e_\mathrm{min}\,,\,E_\mathrm{max}\!=\!e_\mathrm{max})$ be the probability density function (pdf) of the random variables.
We shall also use a shorthand notation $F(E_\mathrm{min},E_\mathrm{max})$, when there is no fear of confusion, and likewise for other random variables.
We denote the $k \times k$ identity matrix by $\mathbb{I}_k$.

The prior distribution of $H$ shall be such that
\begin{multline}
 F(E_\mathrm{min}\!=\!e_\mathrm{min}\,,\,E_\mathrm{max}\!=\!e_\mathrm{max}) \\
 =
  \begin{cases}
 F(\Delta_\mathrm{max}\!=\!e_\mathrm{max}\!-\!e_\mathrm{min})\quad & (e_\mathrm{max} \geq e_\mathrm{min})\\
	0 & (\textrm{otherwise})
  \end{cases},
\end{multline}
where 
\begin{equation}
 \Delta_\mathrm{max} := E_\mathrm{max}-E_\mathrm{min}
\end{equation}
is to be regarded as a random variable.

Each Hamiltonian has $d$ energy eigenvalues $E_1,\dots,E_d$.
We set $E_1 = E_\mathrm{min}$ and $E_d = E_\mathrm{max}$. 
Given a set of specific bounds $(E_\mathrm{min},E_\mathrm{max})\!=\!(e_\mathrm{min},e_\mathrm{max})$, it remains to specify other $d-2$ eigenvalues.
We assume that the conditional density function $F(E_k|E_\mathrm{min},E_\mathrm{max})$ of each eigenvalue $E_k$ is independent of other eigenvalues, i.e., 
\begin{multline} \label{E_k:ind}
F(E_2,\dots,E_{d-1}|E_\mathrm{min},E_\mathrm{max})\\
 = \Pi_{k=2}^{d-1} F(E_k|E_\mathrm{min},E_\mathrm{max}), 
\end{multline}
and each $F(E_k|E_\mathrm{min},E_\mathrm{max})$ is a uniform distribution over the range from $E_\mathrm{min}$ to $E_\mathrm{max}$.

A straightforward argument to obtain a reliable estimate of the upper bound is to choose a value $E_{UB}$ sufficiently large so that {the} probability
\begin{equation}
 p(E_{UB} \geq \Delta_\mathrm{max} \geq 0) = \int_0^{E_{UB}} F(\Delta_\mathrm{max}\!=\!x) dx
\end{equation}
is sufficiently large.
Such $E_{UB}$ is an upper bound for $\Delta_\mathrm{max}$ since $\Delta_\mathrm{max} \leq E_{UB}$.
This estimate, however, may be too conservative, if $F(\Delta_\mathrm{max})$ has a very broad distribution.
A tighter estimation is possible by quantum process tomography, but as already discussed, this requires an exponentially increasing time cost.
The evaluation of $a_{U(t)}$, which we describe next, provides a more efficient {estimation} of $\Delta_\mathrm{max}$.

The estimation protocol on the upper bound of $\Delta_\mathrm{max}$ is as follows:
\begin{enumerate}
 \item Choose numbers $t\!>\!0$ and $\epsilon$ so that $1-\big(\frac{2}{d} + \frac{d-2}{d\pi}\big) \geq \epsilon > 0$.
 \item Set $j\!=\!0$ and $c\!=\!\frac{2}{d}\!+\!\frac{d-2}{d\pi}\!+\!\epsilon$.
 \item Evaluate $a_{U(t/2^j)}$. 
 \item If $a_{U(t/2^j)} < c$, change $j$ as $j+1$ and go back to Step 3, otherwise go to Step 4.
 \item Conclude that $\Delta_{\rm max} \leq 2^j \cdot 2 \pi/t$.
\end{enumerate}
Suppose that the protocol terminates with $j = J$.
The probability $p_{J,\mathrm{fail}}$ that this estimation procedure fails is given by the probability of choosing a Hamiltonian such that satisfies $a_{U(t/2^j)} <c$ for $j$ from $1$ to $J-1$ and $a_{U(t/2^J)} \geq c$, but simultaneously $\Delta_{\rm max} \geq 2^J \cdot 2 \pi/t$, i.e.,
\begin{multline}
 p_{J,\mathrm{fail}} = p\big(\Delta_{\rm max} t/2^J \geq 2 \pi,\\
 a_{U(t)} < c, \dots , a_{U(t/2^{J-1})} <c, a_{U(t/2^J)} \geq c\big),
\end{multline}
where each $a_{U(t/2^j)}$ is used as a random variable.
We will see that
\begin{equation}
 p_{J,\mathrm{fail}}\leq 4 \exp \left( - \frac{3 d \epsilon^2}{6 + 4 \epsilon} \right) \Phi(d,\epsilon), \label{scalebound}
\end{equation}
where
\begin{equation}
 \Phi(d,\epsilon) :=  \left( 1 + \frac{12}{d^2 \epsilon^2 \log (1 + d \epsilon^2 )} \right).
\end{equation} 

\subsection{Evaluation of $a_U$}\label{esta}
\begin{figure}
\begin{center}
	\includegraphics[clip, width=0.7\columnwidth]{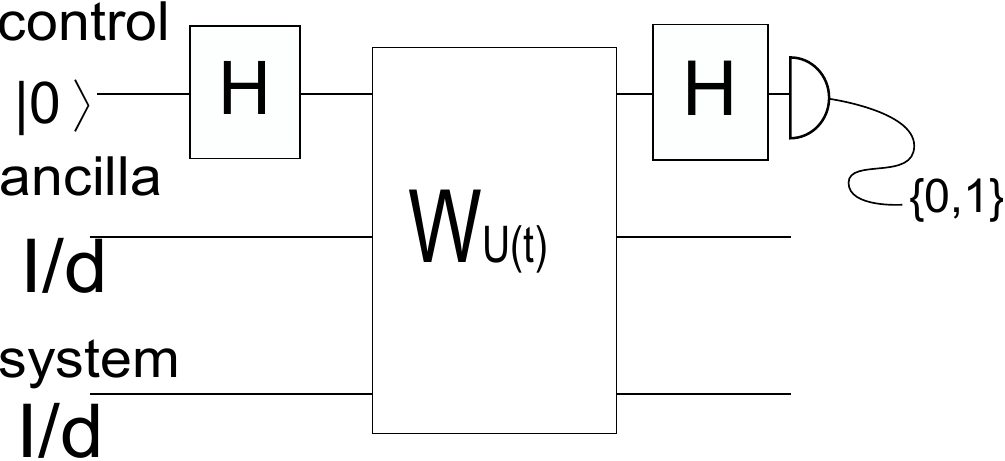}
	\caption{Quantum circuit representation of the algorithm to evaluate the coherence factor $a_{U(t)}$.   $W_{U(t)}$ represents the classically conditioned quantum gate. A quantum circuit representation of $W_{U(t)}$ is given by Fig.\,\ref{pseudo}(a).}
	\label{ADQC1}
\end{center}
\end{figure}

The subroutine algorithm to evaluate $a_U$ is given as follows (Fig.\,\ref{ADQC1}).
First, we prepare a probe consisting of a $d$-dimensional ancilla and control qubit.
The ancilla and target are set to the completely mixed state $\mathbb{I}_d/d$ and the control qubit to the state $\ket{0}$.
The initialization of the target can be achieved, for instance, by first swapping the state of the ancilla with the target, after which we reset the ancilla to $\mathbb{I}_d/d$.

Next, we apply the Hadamard gate $\mathrm{H}$ on the control qubit.
This yields the state
\[
  \ket{+}\bra{+} \otimes \frac{\mathbb{I}_d}{d} \otimes \frac{\mathbb{I}_d}{d},
\]
where the terms in the tensor product correspond to the control qubit, ancilla, and target, respectively. 
We perform the classically conditioned quantum gate $W_U$.
The resulting state is 
\begin{multline}
 \frac{1}{2d^2} \cdot \mathbb{I}_2 \otimes \mathbb{I}_d \otimes \mathbb{I}_d\\
 \hspace{-10mm}  + \frac{1}{4d^2} \cdot \bigg[ \sigma_z \otimes \left(U \otimes U^\dagger + U^\dagger \otimes U\right)\\
  - i\sigma_y \otimes \left(U \otimes U^\dagger - U^\dagger \otimes U\right)\bigg].  
\end{multline}
Another Hadamard gate is applied on the control qubit.
The reduced density matrix of the control qubit becomes
\begin{equation}
\tilde{\rho}_c = \frac{\mathbb{I}_2}{2} + \abs{\tr{U}}^2 \cdot \frac{\sigma_z}{2} =\frac{1}{2}  \left(\mathbb{I}_2 + a_{U(t)}^2 \sigma_z\right).
\end{equation}
Thus the average value obtained from the measurement of $\sigma_z$ on the control qubit is
\begin{equation}
\mathrm{Tr}[ \tilde{\rho}_c \sigma_z] = a_{U(t)}^2.
\end{equation}
Therefore, $M$ iterations of this algorithm estimate $a_{U(t)}^2$ with error of $O\Big(1/\sqrt{M}\Big)$, which decreases independently of dimension. 
Hereafter, we assume that we can obtain $a_{U(t)}$ with sufficiently high accuracy. 

\subsection{Probability of incorrect estimation}\label{probinc}
{In this subsection, we evaluate the failure probability of the energy scale estimation protocol given {by Eq.}~(\ref{scalebound}).}
It is clear that for any two random variables $X$ and $Y$, 
\begin{equation}
 p(X\!\geq\!x , Y\!\geq\!y)~\leq~p(X\!\geq\!x).
\end{equation}
Thus, we have a bound on $p_{J,\mathrm{fail}}$, i.e.,
\begin{equation}
 p_{J,\mathrm{fail}} \leq p\big(\Delta_{\rm max} t/2^J \geq 2 \pi, a_{U(t/2^J)} \geq c\big).
\end{equation}
Thus, it suffices to prove that
\begin{equation}
 p\big(\Delta_{\rm max} t \geq 2 \pi, a_{U(t)} \geq c\big) \leq 4 \exp \left( - \frac{3 d \epsilon^2}{6 + 4 \epsilon} \right) \Phi(d,\epsilon)
\end{equation}
holds for any $t > 0$.

Let us regard $a_{U(t)}$ as a random variable.
Observe that
\begin{multline} \label{a:large-when-DE:large}
 p\big(\Delta_{\rm max} t \geq 2 \pi, a_{U(t)} \geq c\big)\\
 = \int_{2 \pi}^\infty p(a_{U(t)} \geq c' | t\Delta_\mathrm{max}) F(t \Delta_\mathrm{max}\!=\!x)~dx.
\end{multline}
We introduce pdf of $a_{U(t)}$ and $t \Delta_\mathrm{max}$, i.e.,
\begin{equation}
 F\big( a_{U(t)},t \Delta_\mathrm{max} \big).
\end{equation}
For any given $\Delta_\mathrm{max}$, the probability of obtaining $a_{U(t)} \geq c$ for any number $c$ is given by
\begin{equation}
 p\big( a_{U(t)} \geq c|t\Delta_\mathrm{max}\big) = \int_c^\infty F(a_{U(t)}=y|t\Delta_\mathrm{max}) d y,
\end{equation} 
where $F(a_{U(t)}|t\Delta_\mathrm{max})$ is the conditional probability density,
\begin{equation}
 F(a_{U(t)}|t\Delta_\mathrm{max}) = \frac{F(a_{U(t)},t\Delta_\mathrm{max})}{F(t\Delta_\mathrm{max})}.
\end{equation}

We first analyze an average property of $a_{U(t)}$ for a given $t\Delta_\mathrm{max}$.
By definition, $a_{U(t)}$ is the magnitude of a complex number $\gamma_{U(t)} = \Big[ \sum_{k=1}^d \exp(-i E_k t)\Big]/d$.
Recall that the conditional density function $F(E_2,\dots,E_{d-1}|E_\mathrm{min},E_\mathrm{max})$ is given by Eq.\,\eqref{E_k:ind}, where each $E_k$ is a random variable with an independent and identical distribution.
Thus the average $\langle \gamma_{U(t)} \rangle$ is
\begin{equation}
 \langle \gamma_{U(t)} \rangle = \frac{1}{d} \sum_{k=1}^d \langle e^{-i E_k t} \rangle.
\end{equation}
For each $k=2,\dots,d-1$, $F(E_k|E_\mathrm{min},E_\mathrm{max})$ is a uniform distribution over the range between $E_\mathrm{min}$ and $E_\mathrm{max}$, which implies
\begin{equation}
 \langle e^{-i E_k t} \rangle = \frac{e^{-i E_\mathrm{max}t} - e^{-i E_\mathrm{min}t}}{-it \Delta_\mathrm{max}}.
\end{equation}
We see that the r.h.s.\ is independent of $k$.
This shows that 
\begin{equation} \label{boundgamma}
 \abs{\langle \gamma_{U(t)} \rangle} \leq \frac{2}{d} + \frac{d-2}{d} \cdot \frac{2}{t \Delta_\mathrm{max}},
\end{equation}
where we have used the triangle inequality on the numerator and that
\begin{equation}
 \abs{\langle e^{-i E_1 t} \rangle} = \abs{\langle e^{-i E_d t} \rangle} = 1.
\end{equation}

While this only shows an average behavior of $\gamma_{U(t)}$, the following extension of Bernstein's inequality asserts that the average is a good representation of the whole for any sum of independent random variables.
\begin{lem}[Bernstein's inequality (for vectors)\,\cite{vectorbern}] \label{vectorbern}
 Let $\{X_1, \dots, X_d\}$ be a set of $d$ independent (not necessarily identical) random variables that return an $n$-dimensional vector in $\mathbb{C}^n$ and $S$ be $S := \sum_{k=1}^d X_k$.
Given that for each $X_k$, $\norm{X_k - \langle X_k \rangle} \leq 1$,
then for any $\tilde{t} > 0$
\begin{multline} \label{BI}
 p\big( \norm{S - \langle S \rangle} \geq \tilde{t} \big)\\
	 \leq 4 \exp\left( -\frac{\tilde{t}^2/2}{\sigma^2 + (\tilde{t}/3)} \right) \left( 1 + \frac{6}{\tilde{t}^2 \log \big( 1+ (\tilde{t}/\sigma^2) \big)} \right), 
\end{multline}
where $\sigma^2 \equiv \sum_{k=1}^d \langle \norm{X_k}^2 \rangle - \norm{\langle X_k \rangle}^2$.
\end{lem}
\noindent Crudely speaking, it states that the sum of independent random variables distributes most likely around the average of the sum.
This is a particular instance of ``concentration of measure" phenomena known in probability theory\,\cite{com}.

Hence, it is unlikely that $a_{U(t)}$, which is equal to $\abs{\gamma_{U(t)}}$, deviates far from $\abs{\langle \gamma_{U(t)} \rangle}$.
Formally, we arrive at the following bound on conditional probability $p(a_{U(t)} \geq c' | t\Delta_\mathrm{max})$,
\begin{theo} \label{bound-a-given-DE}
 For any $\epsilon > 0$, if $t \Delta_\mathrm{max} \geq 2 \pi$, then
\begin{align}
 &p\left( a_{U(t)} \geq \frac{2}{d} + \frac{d-2}{d\pi} + \epsilon \bigg| t \Delta_\mathrm{max} \right) \nonumber\\
 &\qquad \leq 4 \exp \left( - \frac{3 d \epsilon^2}{6 + 4 \epsilon} \right) \left( 1 + \frac{12}{d^2 \epsilon^2 \log (1 + d \epsilon^2 )} \right).
\end{align} 
\end{theo}
\begin{proof}
By Eq.\,\eqref{boundgamma},
\begin{equation}
 \abs{\langle\gamma_{U(t)}\rangle} \leq \frac{2}{d} + \frac{d-2}{d\pi}.
\end{equation}
Therefore, for any Hamiltonian such that
\begin{equation}
 a_{U(t)} \geq \frac{2}{d} + \frac{d-2}{d\pi} + \epsilon,
\end{equation}
it must be that
\begin{equation}
 \abs{\gamma_{U(t)} - \langle \gamma_{U(t)} \rangle} \geq \epsilon.
\end{equation}
This implies that
\begin{multline}
 p\left( a_{U(t)} \geq \frac{2}{d} + \frac{d-2}{d\pi} + \epsilon \bigg| t \Delta_\mathrm{max} \right) \\
   \leq p\left(\abs{\gamma_{U(t)} - \langle \gamma_{U(t)} \rangle} \geq \epsilon \big| t \Delta_\mathrm{max} \right).
\end{multline}
Next, we apply Bernstein's inequality.
To do so, we regard a complex number as a 2-dimensional vector and take
\begin{align}
 X_k &= \frac{e^{-i E_k t}}{2} \\
    S &= \frac{d \cdot \gamma_{U(t)}}{2} = \sum_{k=1}^d X_k.
\end{align}
Note that pdf for each $k$ satisfies
\begin{equation}
 F\big(X_k\!=\!e^{-i E_k t}/2\big) = F(E_k|E_\mathrm{min},E_\mathrm{max}),
\end{equation}
where we take $X_1$ and $X_d$ as a constant random variable.
It is easy to see that $X_k$ satisfy the necessary conditions to apply Bernstein's inequality.
Let $\tilde{t}$ in Eq.\,\eqref{BI} be
\begin{equation}
 \tilde{t} = \frac{d \epsilon}{2},
\end{equation}
since
\begin{multline}
 p\left( \, \abs{\gamma_{U(t)} - \langle \gamma_{U(t)} \rangle} \geq \epsilon \, \big| \, t \Delta_\mathrm{max} \, \right) \\
  = p\left(\,\norm{S - \langle S \rangle} \geq \frac{d \epsilon}{2}~\bigg|~t \Delta_\mathrm{max} \right).
\end{multline}
Notice that the r.h.s.\ of Eq.\,\eqref{BI} is monotonically increasing with respect to $\sigma^2$,
which satisfies
\begin{multline}
 \sigma^2 \leq \sum_{k=1}^d \langle \norm{X_k}^2 \rangle 
     = \sum_{k=1}^d \bigg\langle \abs{\frac{e^{-i E_k t}}{2}}^2 \bigg\rangle = \frac{d}{4}.
\end{multline}
Therefore, we obtain the desired bound given by
\begin{align}
 &p\left( a_{U(t)} \geq \frac{2}{d} + \frac{d-2}{d\pi} + \epsilon \bigg| t \Delta_\mathrm{max} \right) \nonumber\\
 &\qquad \leq 4 \exp \left( - \frac{3 d \epsilon^2}{6 + 4 \epsilon} \right) \left( 1 + \frac{12}{d^2 \epsilon^2 \log (1 + d \epsilon^2 )} \right).
\end{align}
\end{proof}

\end{document}